\documentclass[11pt]{article}

\usepackage{amsmath}
\usepackage{amssymb}
\usepackage{graphicx}
\usepackage{cite}
\usepackage{color} 
\usepackage{sidecap}

\usepackage{setspace} 
\usepackage[labelfont=bf,labelsep=period,justification=raggedright]{caption}

\date{}

\begin{document}

\begin{flushleft}
{\Large
\textbf{Demography and the age of rare variants.}
}
\\
Iain Mathieson$^{1+*}$, 
Gil McVean$^{1}$ 
\\
{\bf{1}} Wellcome Trust Centre for Human Genetics, University of
Oxford,  UK
\\
$^+$ Current address: Department of Genetics, Harvard Medical School,
Boston, MA, USA
\\
$\ast$ E-mail : iain\_mathieson@hms.harvard.edu
\end{flushleft}

\section*{Abstract}
Large whole-genome sequencing projects have provided access
to much of the rare variation in human populations, which is
highly informative about population structure and recent
demography. Here, we show how the age of rare variants can be
estimated from patterns of haplotype sharing and how these ages can be
related to historical relationships between populations.
We investigate the distribution of the age of 
variants occurring exactly twice ($f_2$ variants) in a
worldwide sample sequenced by the 1000 Genomes Project, revealing
enormous variation across populations. The median age of haplotypes
carrying $f_2$ variants is 50 to 160 generations across populations
within Europe or Asia, and 170 to 320 generations within Africa. Haplotypes shared between
continents are much older with median ages for haplotypes shared
between Europe and Asia ranging from 320 to 670 generations. The distribution of the
ages of $f_2$ haplotypes is informative about their
demography, revealing recent bottlenecks, ancient splits, and more
modern connections between populations. We see the signature of
selection in the observation that functional variants are
significantly younger than nonfunctional variants of the same
frequency. This approach is relatively insensitive to mutation rate
and complements other nonparametric methods for demographic inference.

\section*{Author Summary}
In this paper we describe a method for estimating the age of rare
genetic variants. These ages are highly informative about the extent
and dates of connections between populations. Variants in closely related populations generally arose more
recently than variants of the same frequency in more diverged
populations. Therefore, comparing the ages of variants shared across different
populations allows us to infer the dates of demographic events like
population splits and bottlenecks. We also see that rare functional
variants shared within populations tend to have more recent
origins than nonfunctional variants, which is likely to be the
signature of natural selection.  

\section*{Introduction}

The recent availability of large numbers of fully sequenced human
genomes has allowed, for the first time, detailed investigation of
rare genetic variants. These are highly differentiated
between populations \cite{bustamante2011,nelson2012}, may make an
important contribution to genetic susceptibility to disease
\cite{nejentsev2009,johansen2010,mcclellan2010,rivas2011,beaudoin2013},
and provide information about both demographic history, and fine-scale
population structure \cite{gravel2011,mathieson2012}. While patterns
of rare variant sharing are informative in themselves, knowing
the age of the variants allows us to observe changes in structure over
time, and thus to infer the dates of demographic events.

Rare variants are typically more recent than common variants 
and in fact, the age of a variant can be estimated directly from its frequency
\cite{kimura1973,griffiths1998,fu2012}. However there are two problems
with this approach. First, using only the frequency information means
that we cannot distinguish differences between the ages of variants
which are at the same frequency which, as we demonstrate here, can be both large and
important. Second, in order to use this approach, we have to know
the demographic history of the populations involved. In this article, we describe an
alternative approach which uses the fact that the lengths of shared
haplotypes around variants are informative about their ages
\cite{palamara2012,ralph2013,harris2012}. 

Specifically, we estimate the time to the most recent common ancestor
(TMRCA) for $f_2$ haplotypes, which are regions where two chromosomes 
are each other's unique closest relative in a sample. To find these
regions, we look for variants which occur exactly twice in the
sample ($f_2$ variants, or doubletons). We then use nearby variation
to estimate the extent of the $f_2$ haplotype and use
the length of, and number of mutations on, this haplotype to infer its age,
and therefore a lower bound for the age of the variant.  Every $f_2$ variant identifies
an $f_2$ haplotype, but we do not detect all $f_2$ haplotypes because
not all of them carry mutations.  This approach
is fast, robust, and finds shared haplotypes directly from
genotype data, which avoids the need for statistical phasing. We apply this method to the
1000 Genomes phase 1 dataset \cite{1000genomes2012}, 
to quantify the distribution of the ages of variants
shared within and between populations, and between variants in
different functional classes. We demonstrate dramatic differences
between the ages of variants shared across different populations, 
and reveal the signatures of both demography and selective constraint. 

\section*{Results}

\begin{figure}[]
\begin{center}
\includegraphics{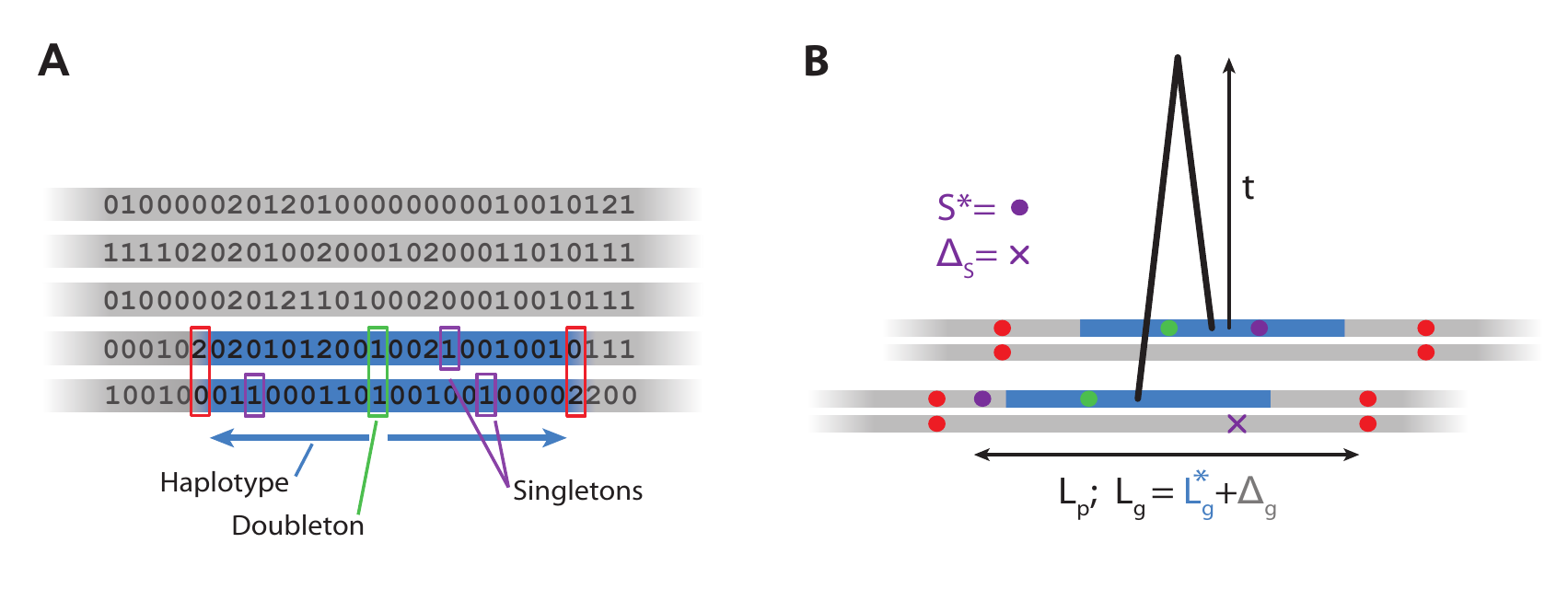}
\end{center}
\caption{
{\bf{Algorithm and model for haplotypes.}} {\bf{A}}:
Algorithm for detecting $f_2$ haplotypes. For each $f_2$ variant
in the sample (green), we scan left and right until we find
inconsistent homozygote genotypes (red), record the physical and
genetic length of this region (blue), and the number of singletons
(purple). {\bf{B}}: Model for haplotype age $t$. Consider the 4
chromosomes (grey) of the two individuals sharing an $f_2$ haplotype
(blue). We model the total genetic length of the inferred haplotype, $L_g$, as the sum of the true
genetic length $L_g^*$ and an error $\Delta_g$. Similarly, we model
the number of singletons $S$ as the sum of the number on the shared
chromosome ($S^*$) and the number on the unshared chromosomes, 
$\Delta_S$. We ignore the fact that we overestimate $L_p$ and
therefore that some of the singletons might lie in the unshared part
of the chromosome. 
}
\label{Fig1}
\end{figure}

We first give a brief outline of our approach (Figures \ref{Fig1}, S\ref{FigS1},
{\bf{Methods}}). Given a sample of individual
genotypes, we find all $f_2$ variants. That is, variants which
have exactly two copies (in different individuals) in the sample. 
This tells us that, in the absence of repeat mutations and assuming
that the $f_2$ variant is derived, those individuals must share an
$f_2$ haplotype at that
position. We then scan left and right along the genome, until we reach
a point where the two individuals have inconsistent homozygote
genotypes (0 and 2, Figure \ref{Fig1}A).

Using both the genetic and physical lengths of the region, and the
number of singletons, we compute an approximate likelihood for the age
of the haplotype (Figure \ref{Fig1}B). We use the data to estimate error
terms to take into account the fact that the algorithm described above
does not find the shared haplotypes precisely. Then, for each
haplotype, we find the maximum likelihood estimate (MLE) of the age of
each haplotype. We investigate the distribution of these MLEs for
different classes of $f_2$ variants, for example those shared within
or between specific populations. 

\subsection*{Simulation results}

\begin{figure}
\begin{center}
\includegraphics{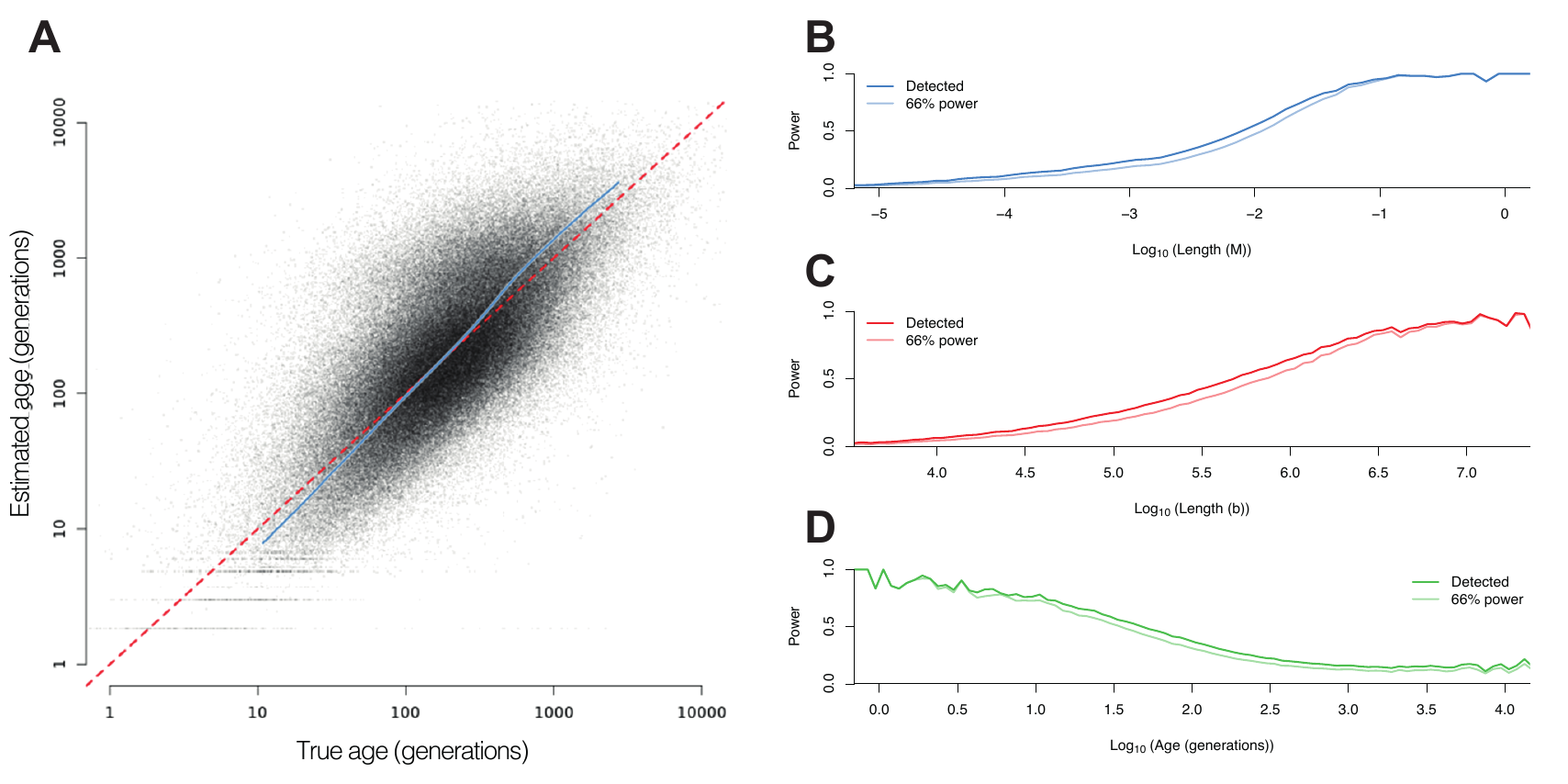}
\end{center}
\caption{
{\bf{Estimating $f_2$ age from simulated data.}}
We simulated whole genomes
for 100 individuals (200 chromosomes), with $N_e=14,000$,
$\mu=1.2\times10^{-8}$ and HapMap 2 recombination rates. {\bf{A}}: Estimated
age against true age. The grey dots are the MLEs for each detected
haplotype. The blue line is a quantile-quantile (qq) plot for the MLEs
(from the 1$^{st}$ to 99$^{th}$ percentile).  {\bf{B}}-{\bf{D}} Power to detect
$f_2$ haplotypes as a function of {\bf{B}}: genetic length, {\bf{C}}:
physical length and {\bf{D}}: haplotype age; in each case the darker
line represents the power to detect $f_2$ haplotype with 100\% power
to detect $f_2$ variants, and the lighter line the power with 66\%
power to detect variants. 
}
\label{Fig2}
\end{figure}

To test our approach, we ran whole genome simulations for a sample of
100 diploid individuals with MaCS\cite{chen2009a}, using the combined
HapMap 2 recombination maps \cite{hapmap2}, and a mutation rate
($\mu$) of $1.2\times 10^{-8}$ per-base per-generation, assuming a constant
effective population size ($N_e$) of 14,000; chosen to reflect
parameters relevant to human genetic variation. We investigated both our power to detect the $f_2$
haplotypes and how accurately we could estimate
the distribution of $f_2$ ages (Figure \ref{Fig2}). We detected around 26\% of all $f_2$
haplotypes. Unsurprisingly, we have more power to detect very long
haplotypes, but we detected many small haplotypes as well: 19\% of our
total had true genetic length less than than 0.1cM. Having imperfect
power to detect $f_2$ variants does not have a large effect on our power
to detect $f_2$ haplotypes since most haplotypes carry more than one
$f_2$ variant. We have higher power for more recent haplotypes because
they are longer but, at least for a population of constant size, 
 this effect is cancelled to some extent
for older haplotypes because the branches above them tend to be longer
and therefore more likely to carry mutations. 

There is high uncertainly in the age of any individual haplotype 
(Figure \ref{Fig2}A). However, we can
compute well-calibrated confidence intervals (Figure S\ref{FigS2}). In this
example, the median MLE of the age of the detected haplotypes is 179
generations and the true median is 192 generations. The median width
of the 95\% confidence interval is 730 generations. Information about the ages comes
mainly from the genetic length, and the principal
advantage of the singleton information is for very old haplotypes
where the length-based estimator is otherwise biased (Figure S\ref{FigS3}).

In addition, we ran simulations to check that the model was robust to
more complicated demographic scenarios including splits, bottlenecks
and expansions, as well as mis-specification of $N_e$ (Figure S\ref{FigS4}).
We also investigated the effect that these scenarios had
on the distribution of the ages of the $f_2$ haplotypes, demonstrating
that we could detect the signatures of demographic events. For example,
population bottlenecks lead to a high density of $f_2$ haplotypes
during the bottleneck and, following a population split haplotypes
shared between populations have median age roughly equal to the split
time (Figure S\ref{FigS5}).

\subsection*{1000 Genomes data}

We applied our estimator to the phase 1 data release of the 1000
Genomes Project \cite{1000genomes2012}, which consists of whole-genome
variant calls for 1,092 individuals drawn from 14 populations (Table
\ref{Tab1}). We used two of the 1000 Genomes callsets;  
one made from sequence data, and one made using a dense genotyping array. 
Restricting our analysis to the autosomes,
we extracted $f_2$ variants from the sequence data callset,
and then detected haplotype lengths around them (that is, the distances to
incompatible homozygotes), using only the array data, to minimise the effect of genotyping
errors. We then counted $f_1$ variants on these haplotypes from the
sequence data callset. From 4,066,530 $f_2$ variants we detected 1,893,391 $f_2$ haplotypes, with
median genetic and physical lengths of 0.7cM and 600kb
respectively. The median number of singletons per haplotype was 3. Of the 1.9
million $f_2$ haplotypes, 0.7 million were shared within populations
and 1.5 million were shared within continents. Sharing 
of $f_2$ variants largely reflects expected patterns of
relatedness on a population level, and also reveals substructure in
some populations, notably GBR (Figure S\ref{FigS6}). 

\begin{table}[]
\begin{tabular}{r|ll}
Abbreviation & Sample size & Description\\
\hline
ASW & 61 &African Ancestry in SW USA\\
LWK & 97 & Luhya in Webuye, Kenya\\
YRI & 88 & Yoruba in Ibadan, Nigera\\
CLM & 60 & Colombians in Medell\'{i}n, Colombia\\
MXL & 66 & Mexican Ancestry in Los Angeles, CA, USA\\
PUR & 55 & Puerto Ricans in Puerto Rico\\
CHB & 97 & Han Chinese in Beijing, China\\
CHS & 100 & Han Chinese South\\
JPT & 89 & Japanese in Tokyo, Japan\\
CEU & 85 & Utah residents with ancestry from northern and western
Europe\\
FIN & 93 & Finnish in Finland\\
GBR & 89 & British from England and Scotland\\
IBS & 14 & Iberian Populations in Spain\\
TSI & 98 & Toscani in Italy\\
\end{tabular}
\caption{Short descriptions of the 1000 Genomes populations.}
\label{Tab1}
\end{table}

\begin{figure}
\begin{center}
\includegraphics[width=\textwidth]{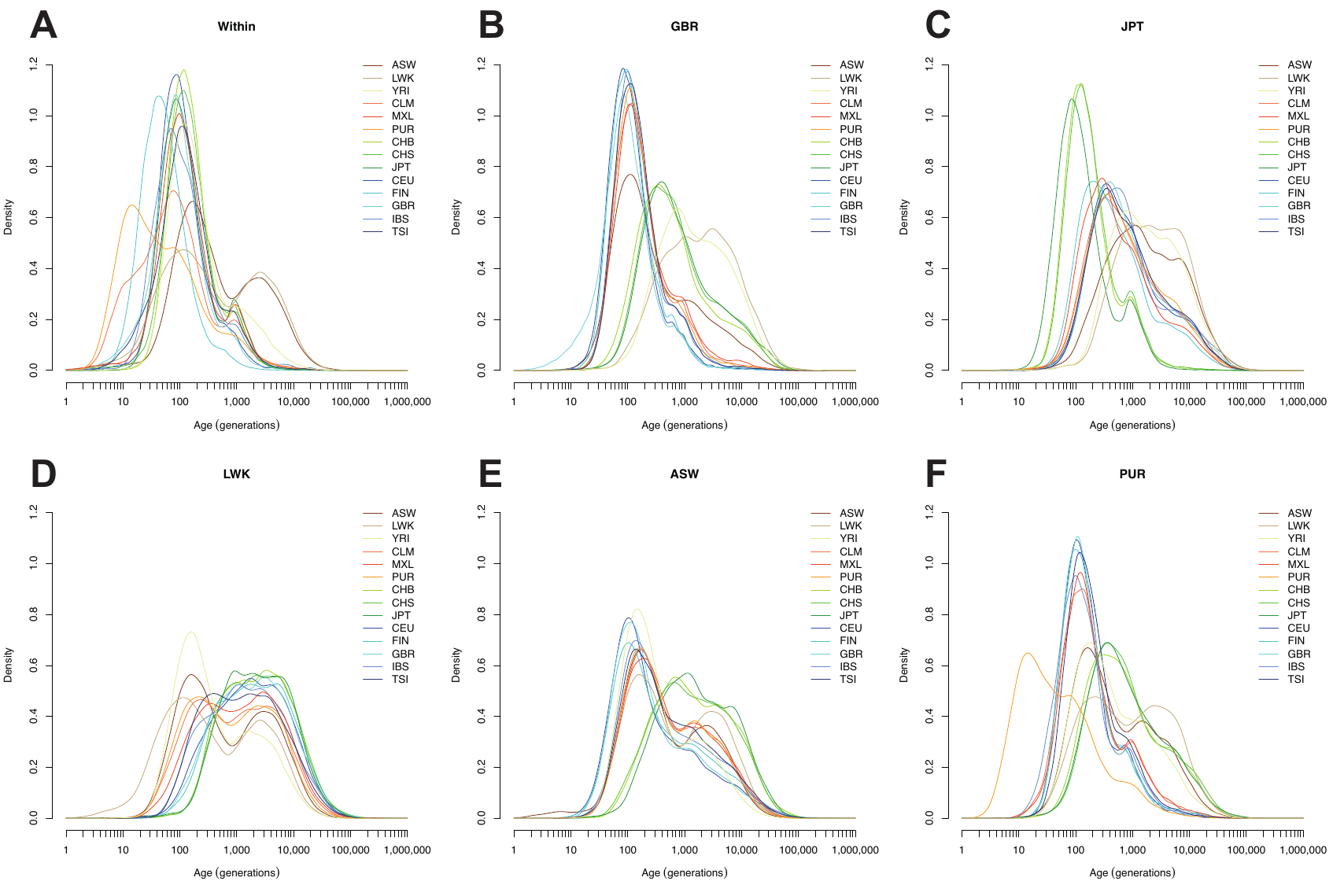}
\end{center}
\caption{
{\bf{The estimated age distribution of $f_2$ haplotypes.}} {\bf{A}}: The
distribution of the MLE of the ages of haplotypes shared within each
population. {\bf{B}}-{\bf{F}}: The distribution of the MLE of the ages of haplotypes shared
between one population and all other populations, shown for each of GBR, JPT,
LWK, ASW, and PUR. Populations are described in Table \ref{Tab1}.
Density estimates are computed in $\log_{10}$ space, using the
base R ``density'' function with a Gaussian kernel. 
}
\label{Fig3}
\end{figure}

We used the combined recombination rate map from HapMap 2 to determine
genetic lengths, and assumed a mutation rate of $0.4\times10^{-8}$
per-base per-generation (reflecting a true mutation rate of $1.2\times
10^{-8}$ multiplied by a power of $\frac{1}{3}$ to detect
singletons \cite{scally2012,1000genomes2012}). We estimated $N_e=185,000$ (from
the number of singletons in the dataset, {\bf{Methods}}). We then computed
MLEs for the ages of all the $f_2$ haplotypes shared
between every pair of populations (Figures \ref{Fig3}, S\ref{FigS7},
Tables S\ref{TabS13}-S\ref{TabS17}).
We estimated that on most chromosomes the
median overestimate in the haplotype lengths is 0.1-0.15cM (but
more on chromosomes 1, 9 and 15), and that $\theta$
(estimated from singletons) was around $3.7\times10^{-3}$ and
$2.4\times10^{-3}$ per-base for African and Non-African populations
respectively (Table S\ref{TabS18}). 

For haplotypes shared within populations (Figure \ref{Fig3}A), the MLEs of
haplotypes within most European and Asian
populations are clustered around 100 generations ago. For
example, the median age of GBR-GBR haplotypes is 90 generations. 
PUR and, to a lesser extent, CLM have many
very recent haplotypes (peaking around 11 generations
ago), consistent with a historical bottleneck in these
populations 300-350 years ago. FIN haplotypes peak around 14
generations (400-450 years) ago. African populations have many recent haplotypes but
also a much longer tail than the other populations, with 
ancestry apparently extending back for thousands of generations.
For example the median age of LWK-LWK haplotypes is
320 generations, but the 95\% quantile is 8,500 generations.

Between-population sharing is largely consistent with
 the historical relationships among populations (Figure \ref{Fig3}B-D).
 Within continents, sharing within Asia or Europe has a median
of 50-160 generations, depending on the populations, 
and sharing within Africa 170-340 generations. Sharing between
continents is much older, with median Asian-European sharing 320-670 generations
old, and Asian-African sharing rather older, with a median around
2,300 generations ago for LWK and 1,700 generations ago for YRI.
The age of European-African sharing varies between
populations, from 1,000 to 2,000 generations ago, but is 
more recent than Asian-African sharing, perhaps
suggesting greater subsequent migration between these continents.  We
discuss these figures in the context of split times and migrations in
the {\bf{Discussion}}. 

Admixed populations have age distributions that are combinations of
the distributions of the admixing populations (Figure \ref{Fig3}E-F).
 Even in these populations we can see signs of more subtle
history. For example, GBR-CLM haplotypes have an age distribution
which looks more like GBR-TSI or GBR-IBS than GBR-CEU, presumably representing the
fact that the major contribution to European admixture in the Americas
is from southern Europe (Figure S\ref{FigS8}).  

\begin{SCfigure}
\includegraphics[width=8.3cm]{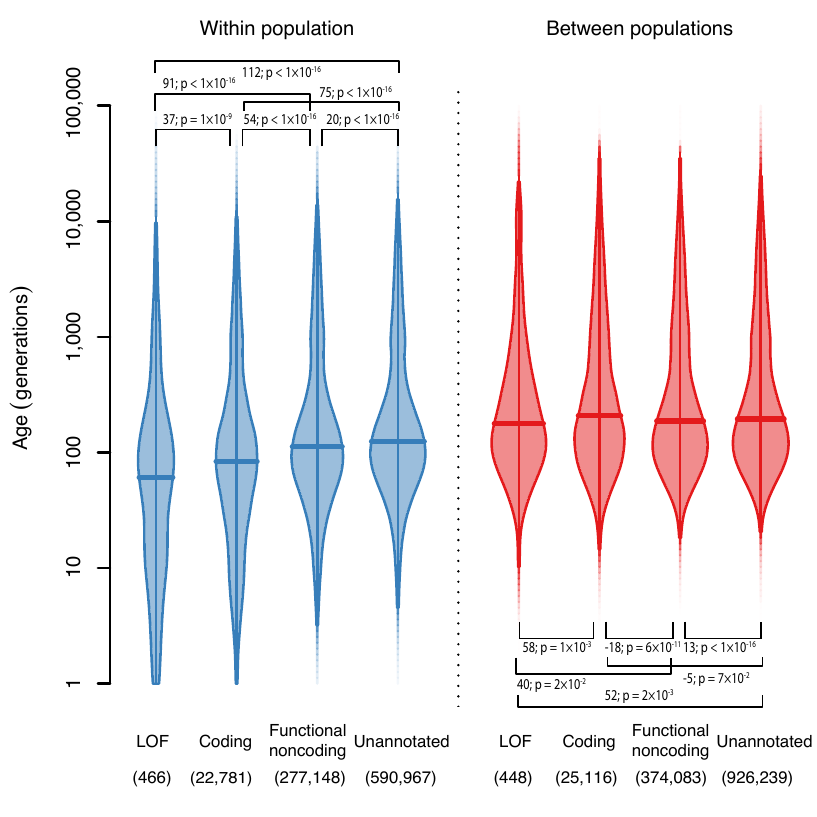}
\caption{
{\bf{The ages of haplotypes around $f_2$ variants with
different functional annotations.}} Density is indicated by the width of the
shape, and horizontal bars show the median.
 We show separately the densities for $f_2$ variants
shared within a population (left, blue), and $f_2$ variants shared
between populations (right, red). Numbers in brackets show the number
of variants in each class. Bars show the pairwise differences in
means, and $t$ test p-values for a difference in log means between groups. 
}
\label{Fig4}
\end{SCfigure}

We also looked at the distribution of the ages of $f_2$ variants
broken down by functional annotation (Figure \ref{Fig4},
{\bf{Methods}}). We found that for variants shared within a single
population, loss-of-function (LOF) variants are younger than coding
variants, which are younger than functional noncoding variants, and
all annotated variants are younger than unannotated variants. The
median ages of these variants are 58, 83, 112 and 125 generations for
LOF, coding, functional noncoding and unannotated variants
respectively. This is presumably because purifying selection against
damaging mutations means that functional variants are less likely to
become old (though positive selection for beneficial mutations would
have the same effect). This effect has previously been both predicted
and observed \cite{maruyama1974,kiezun2013}. However, it is not
strictly true for variants shared between different populations and, in fact, the
effect is partially reversed (median ages are 176, 205, 186 and 195
generations for LOF, coding, functional noncoding and unannotated
variants).  A likely explanation is that functional variants surviving
long enough to be shared between populations are selectively
neutral or recessive and thus unaffected by selection at low
frequency. This suggests that studies looking for disease-causing rare
variants should concentrate on variants private to a single
population, since variants shared across populations are unlikely to
have large phenotypic effects.

\subsection*{Robustness}

This analysis requires us to estimate several parameters, and in this
section, we investigate how robust it is to varying them.

The parameter $k$ is related to the probability of
discovering $f_2$ haplotypes. We know that $1\leq k \leq2$. $k=1$ implies that the
probability that we discover a haplotype is independent of its length
while $k=2$ implies that this probability increases linearly with
length. We chose $k=1.5$ based on simulations, but it may be the case that this is not the
optimal $k$ for real data. To test how much of an impact this might have, we
 re-ran the analysis of the 1000 Genomes data using $k=1$ and $k=2$.
Larger values of $k$ increase our age estimates. For example, the median
CEU-CHB age is 403, 481 and 560 generations using $k=1$, 1.5 and 2
respectively. Overall, setting $k=2$ increases the median age
estimates by between 6 and 30\%, depending on population, 
with more recent ages more sensitive to $k$.  

The parameters $k_e$ and $\lambda_e$ are the shape and
rate of the (gamma) distribution of the overestimate of haplotype lengths
({\bf{Methods}}). We estimated these parameters separately from the
array data for each
chromosome (Table S\ref{TabS18}). We noticed that chromosomes 1, 5
and 9 had estimated parameters that implied a greater overestimate
(larger $k_e$, smaller $\lambda_e$), presumably due to the density of
markers on the array for those chromosomes. 
In addition, these chromosomes had older estimated haplotype
ages, for example we estimated that the median age of $f_2$ haplotypes
on chromosome 1 was 16\% higher than the median age of haplotypes on
chromosome 2, suggesting that our error model is not fully robust to variation
in marker density.

\section*{Discussion}

We described an approach to estimate the age of $f_2$
haplotypes, without making any prior assumptions about population
structure or history. Though the age of any individual haplotype is uncertain,
major features of the distribution of haplotype ages are detected, 
demonstrating qualitative differences between populations that are
almost certainly due to past demographic events. The next important
question is to what extent we can use these distributions as quantitative
estimates of the ages of demographic events. 

\begin{figure}
\begin{center}
\vskip-1cm
\includegraphics[width=17.35cm]{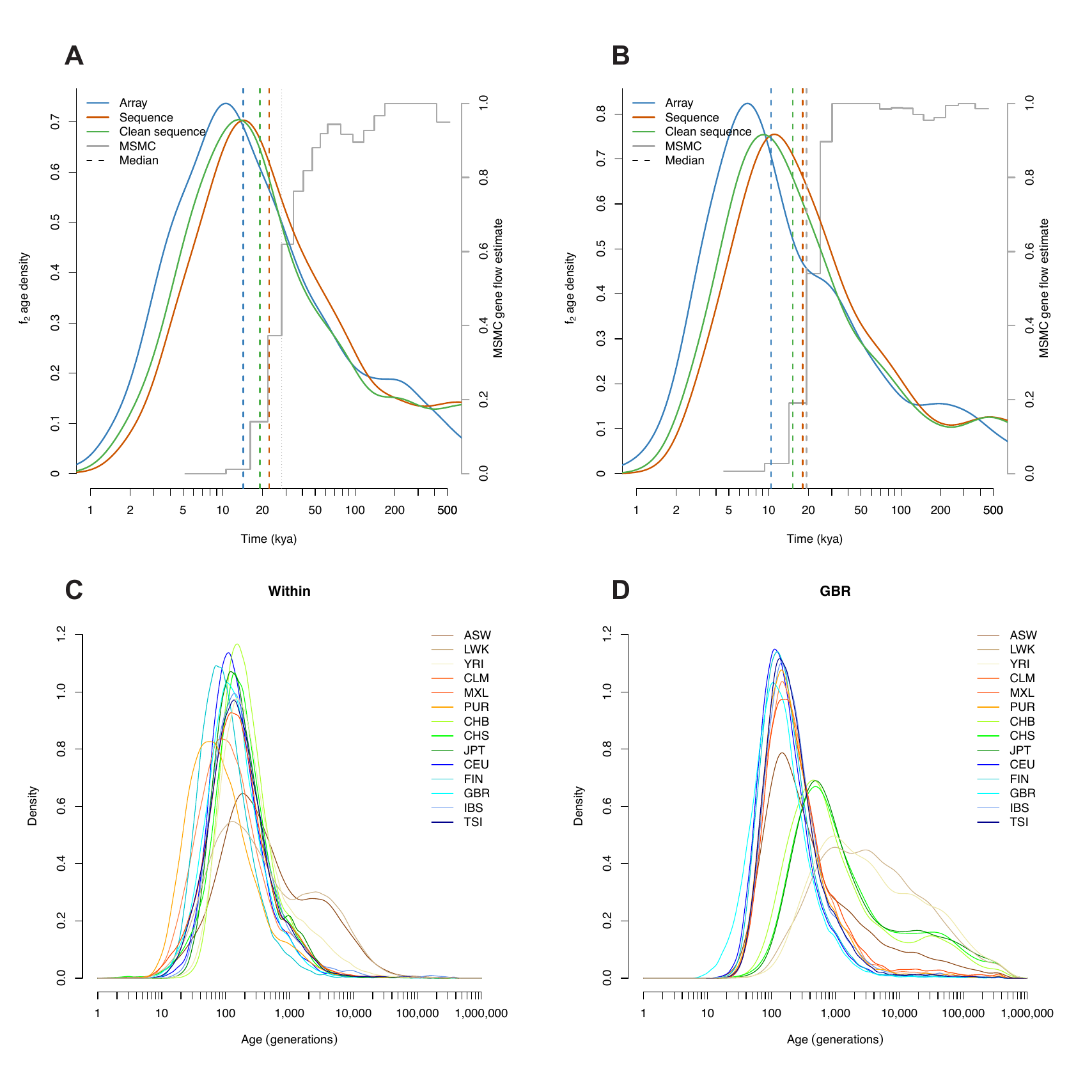}
\end{center}
\vskip-1cm
\caption{
{\bf{Comparison with MSMC, and the effect of estimating haplotypes with
sequence data.}} {\bf{A}}: The age distribution of $f_2$ haplotypes
shared between CHB and CEU estimated with array, sequence and 
``clean'' sequence (with indels and low
complexity regions removed; {\bf{Methods}}). Coloured dashed lines show
the medians of each distribution. The grey stepped line shows
relative cross-population coalescence rates estimated by MSMC (S. Schiffels,
personal communication), and the grey dashed line shows the earliest
point where this rate is less that 0.5. In both cases, we assume 30 years per
generation and $\mu=1.25\times 10^{-8}$.  {\bf{B}}: As in {\bf{A}} but
for $f_2$ haplotypes shared between CHB and MXL, restricted to
haplotypes where the MXL individual is inferred to be homozygous for
Native American ancestry. {\bf{C-D}}: Age
distributions inferred using ``clean'' sequence data, comparable to
Figure 3A-B (Note the extended x-axis).
}
\label{Fig5}
\end{figure}

Consider the split between European and East Asian
populations. Model based estimates of this split time have
ranged from 14 to 40 thousand years ago (kya)  \cite{keinan2007,gutenkunst2009,gronau2011}.
Although, these are likely to be too low because they
assumed a mutation rate, $\mu$, of $2-2.5\times10^{-8}$ per-base per-generation, 
now thought to be an overestimate
\cite{scally2012} and so a more reasonable range of estimates might be 22-80kya.
The nonparametric PSMC approach\cite{li2011}
estimated a split time of around 22kya (if a lower mutation rate of
$1.25\times 10^{-8}$ is used, 11kya with the higher rate),
and a similar method, MSMC, estimates a split time of 20-40kya
(S. Schiffels, personal communication; Figure \ref{Fig5}A).
Simulations suggest that, under a clean split model, the median of our
estimated ages is close to or slightly below the split time, at least
for recent splits (less than 1,000 generations; Figures S\ref{FigS5} and S\ref{FigS9}).
Comparing CEU to each of CHB, CHS and JPT, taking the median of our
haplotype ages, and assuming a generation time of 30
years\cite{fenner2005}, would imply split times of 14, 17 and 18kya
respectively. Other European populations give different estimates, but mostly between 15
and 20kya.

Similarly, when we looked at $f_2$ variants shared between East Asia
and America (CHB-MXL, but restricting to regions homozygous for
Native American ancestry in MXL; {\bf{Methods}}), we found that the
median age was around 10kya, substantially more recent than the split
time estimated using MSMC (S. Schiffels, personal communication;
Figure \ref{Fig5}B). This seems low, given geological evidence that the
Bering land bridge was submerged by 11-13kya, although a seasonal or 
maritime route likely remained open after that time 
\cite{brigham-grette2004,keigwin2006,meltzer2009}.

Our dates are all around or below the low end of published
estimates, even after we take into account the fact that the median
might be lower than the split time (we estimate about 11\% lower for a
500-generation old split; Figure S\ref{FigS5}D).
There are  several non-exclusive explanations for
this observation. First, post-split gene flow could explain the
discrepancy.  As we have greater power to detect $f_2$ haplotypes if they are more
recent, when the split is not clean many of the haplotypes we observe will derive from the
post-split gene flow rather than from before the initial
split  (Figure S\ref{FigS9}B). In this scenario, we would be detecting the most
recent haplotypes, and our dates would be closer to the most recent
date of contact, rather than the initial split date. 

An alternative explanation might be systematic errors in our estimates.
As we described in the {\bf{Results}}, the approach
is sensitive to the estimated parameters $k$, $k_e$ and
$\lambda_e$. At the extreme, increasing $k$ from 1.5 to its maximum value of 2 would increase
the median age of CEU-CHB haplotypes from 14kya to 17kya. 
To investigate sensitivity to $k_e$ and $\lambda_e$, 
we repeated the analysis, but using sequence data rather than
array data to find the length of the haplotypes (Figure \ref{Fig5}). We note that
when we estimated $k_e$ and $\lambda_e$ using sequence
data they vary very little across chromosomes (Table S\ref{TabS18}). The ages
estimated using sequence data were older than those estimated using
array data (Median age of CEU-CHB haplotypes 23kya, Figure \ref{Fig5}A-B). 
We might expect that sequence data, being more dense
than array data, would find haplotypes more accurately. 
However we would also expect that genotype errors, more
common in sequence than array data,  would make all haplotypes look 
older, by incorrectly breaking haplotypes. Removing indels and low
complexity regions (LCRs; {\bf{Methods}}) thought to be enriched for
genotyping errors from the sequence data reduced the difference
(median CEU-CHB age of 19kya), suggesting 
that around half the increase in age is driven by 
errors. Further, the haplotype ages estimated from sequence data do
not contain the very young (long) haplotypes within CLM, FIN and PUR, which we
independently believe to be correct (Figure \ref{Fig5}C), and also contain a
long tail of extremely old haplotypes which seems unlikely (Figure
\ref{Fig5}D). 

Another source of systematic errors could be the use of incorrect
mutation or recombination rates.  There is
considerable uncertainty about the mutation rate in humans, but 
our approach is relatively insensitive to this, so if the true rate is higher than
$\mu=1.25\times 10^{-8}$ per-base per-generation then mutational clock
based methods which scale linearly with mutation rate
will overestimate the ages of events, thus reducing the discrepancy.

On the other hand, our approach might be
sensitive to errors in the recombination map. We tested
this by running simulations with a different
genetic map to the HapMap map that we used to determine genetic
length. We tested a population-based African
American map\cite{hinch2011}, a map derived from an Icelandic pedigree
\cite{kong2002} and a chimpanzee map from a small
population\cite{auton2012}, but none of these made a substantial
difference to the results and we conclude that the length of the
haplotypes we investigate is sufficiently large that they are robust
to the uncertainty in the recombination map (Figure S\ref{FigS10}).

Finally, systematic errors might occur due to homoplasy
(where the same mutation occurs independently on two
different lineages). While this rate is expected to be low, it may be
locally high in some parts of the genome, for example in CpG islands
which have an order of magnitude higher mutation rate than the genomic
background. If such false positives do occur, they would appear as very
short haplotypes that we would infer to be very old, so they cannot
explain our systematically lower ages. On the other hand, it is likely
that some of the very old haplotypes we see are, in fact, due to
repeat mutations and, in particular, this might explain some of the
very old haplotypes discovered with sequence data.

However, while systematic biases in our estimates might explain some
of the difference between our estimated ages and independent split
time estimates, they cannot explain the
observation that the age distributions vary greatly between
different pairs of populations. This strongly suggests that there is
variation in the extent of gene flow. For example, Asian-FIN sharing
 seems to be more recent than other Asian-European sharing, suggesting relatively recent
contact between East Asian and Finnish populations, compared to other
European populations. It seems likely that worldwide demographic
history is sufficiently complicated that trying to estimate a single
Asian-European (or African-Non African) split time is futile, 
and that a complex model of many splits, migrations and admixtures 
is required to explain the relationship between different populations. 

Ultimately, we would like to be able to make explicit estimates of
parameters like historical effective population size, and the dates of
demographic events. Though the approach we describe here is limited in
in this respect, there is a clear path to extend it to do so.
We could first use a similar approach to estimate the
age of variants at frequency three and higher. Then, treating the
estimates of haplotype ages as estimates of coalescent times, we could
use the empirical distribution of coalescent times to estimate
population sizes and cross-population migration rates as a function of
time. Another improvement would be to use information from the full likelihood surface for each
haplotype, rather than just the point estimate of the age as we do
here. Since, for large samples, we would have good estimates of recent
coalescent rates, we expect that this approach would be very accurate
at inferring recent history, making it a complementary approach to
sequential Markovian coalescent based methods which are typically accurate in the ancient past,
but less so for very recent history. 

\section*{Methods}
\subsection*{Definitions}
Suppose we have a sample of size $N$ of genotypes from a
single genetic region. Define
an $f_2$ variant to be one which occurs exactly twice in the sample in
different individuals. That is, either two individuals have genotype 1
and all the others have genotype 0, or two individuals have genotype 1 and
the others have genotype 2. We assume that the minor allele is
the derived allele. Under the neutral coalescent, for a sample of $2N$
chromosomes, an $f_2$ minor allele will be the derived allele with probability
$\frac{2N-1}{2N+1}\approx 1$ for large $N$ so this is a reasonable assumption for large
samples. 

Define an $f_2$ haplotype shared between chromosomes $a$ and $b$ to be
a region satisfying the following two conditions: 1) The time to the most
recent common ancestor (TMRCA) of $a$ and $b$ does not change over the
region. 2) At one or more sites in the region, $a$ and $b$ coalesce
with each other before either of them coalesce with any other
chromosome. In other words, they are unique genealogical nearest
neighbours (Figure S1). We call the TMRCA of
$a$ and $b$ the age of the haplotype. Additionally, we say that individuals $i$ and $j$ ($i\neq
j$) share an $f_2$ haplotype if $a$ is one of $i$'s two chromosomes and $b$ is one
of $j$'s two chromosomes. 

The problem we solve is to find the $f_2$ haplotypes and then 
estimate their ages. Since each $f_2$ variant must lie in an $f_2$ haplotype, the variants provide a
simple way of detecting the haplotypes. We use the algorithm described
in the main text to find regions which should be larger
 than the $f_2$ haplotypes. The next problem is to determine the
likelihood of the age. We describe our approximate likelihood below
but first, as an example, we describe exact inference in the absence of
confounding factors.  

\subsection*{Exact case}
Suppose we knew the exact genetic and physical lengths of an
$f_2$ haplotype and the number of singletons it carries. Call these
quantities $L_g^*,L_p^*$ and $S^*$. Let the age of this 
haplotype be $t$ generations, or $\tau$ in coalescent time
($\tau=\frac{t}{2N_e}$). Then, for a randomly chosen $f_2$ haplotype
(but not a haplotype at a randomly chosen position, discussed in the
next section), 
$L_g^*$ has an exponential distribution with parameter $4N_e\tau$ and
$S^*$ has a Poisson distribution with parameter $\theta L_p^* \tau$
where $\theta=4N_e\mu$ and $\mu$ is the per-base per-generation
mutation rate. Therefore (ignoring terms that do not depend
on $\tau$), the
log-likelihood of $\tau$ given $L_g^*,L_p^*$ and $S^*$ is 
\begin{equation*}
\ell \left( \tau ; l_g^*,l_p^*,s^*\right)=\left(1+s^*\right)\log\left(\tau\right)-4N_e\tau
l_g^* -\theta l_p^* \tau
\end{equation*}
and the maximum likelihood estimator of $t$ is therefore
\begin{equation*}
\hat{t}=\frac{1+s^*}{2\left( l_g^*+\mu l_p^*\right)}.
\end{equation*}

\subsection*{Approximate likelihood for genetic length}

There are two corrections to the likelihood for genetic length. The
first relates to the ascertainment process of the haplotypes, and the
second to the overestimate in the length due to the way we detect the
endpoints. 

The ascertainment problem is as follows. Suppose we pick a haplotype
at random, then its length is exponentially distributed (i.e. gamma
with shape parameter 1). However, if we pick a point on the sequence
at random then the distribution of the length of the haplotype in
which it falls is gamma distributed with shape parameter 2. This is an
example of the ``inspection paradox'' and it is because in
the second case, we are sampling haplotypes effectively weighted by
their length. In our case, we detect haplotypes if they contain one or
more $f_2$ variants. Therefore the probability that we find a
haplotype is increasing with its physical length (because longer haplotypes are
more likely to carry $f_2$ variants), but sub-linearly. The probability
also increases with genetic length, but in a complex way that
depends on the variation of recombination and mutation rate along the
genome, the age of the haplotype and the demographic history of the
population. For example, in a constant sized population, older
 haplotypes are likely to have longer branches 
above them, and therefore to have more $f_2$ variants, but in an expanding
population the opposite may be true. Rather than trying to take all of these effects into
account, we made the simplifying assumption that we could model the
genetic length $L_g^*$ as a gamma distribution with shape parameter $k$ where
$1<k<2$ and rate $4N_e\tau$. Simulations suggested that $k$ around 1.5
was optimal (Figure S11), and we used this value
throughout. 

The second correction involves the overestimate of
genetic length. We tried to detect the ends of the haplotype by
looking for inconsistent homozygote genotypes, but of course in
practice, after the end of the $f_2$ haplotype, there will be some
distance before reaching such a site. This (genetic) distance $\Delta_g$ is the
amount by which we overestimate the length of the haplotype. We
estimate the distribution of $\Delta_g$ for a given sample by sampling
pairs of genotype vectors, then sampling sites at random and computing
the sum of genetic distance to the first inconsistent homozygote site
on either side. We then fit a gamma distribution with (shape, rate)
parameters $(k_e, \lambda_e)$ to this distribution, for each
chromosome. The likelihood of
$\tau$ is given by the convolution density of $L_g^*$ and
$\Delta_g$, 
\begin{equation}
L(\tau;l_g)=\int_0^{l_g} f_{\gamma}\left(x; \left(k,
    4N_e\tau\right)\right) f_{\gamma}\left(l_g-x; \left(k_e,
    \lambda_e\right)\right) dx
\label{ll_Lg}
\end{equation} 
where $f_{\gamma}\left(x;
  (\kappa,\lambda)\right)=\frac{1}{\Gamma(\kappa)}\lambda^\kappa x^{\kappa-1} e^{-\lambda x}$ is
the density of a gamma distribution with (shape, rate) parameters $(\kappa,
\lambda)$. This integral, and therefore the loglikelihood
$\ell(\tau;l_g)=\log\left[L(\tau;l_g)\right]$ can be expressed in
terms of the confluent hypergeometric function $\,_1F_1$ (ignoring
terms that do not depend on $\tau$),

\begin{equation}
\ell(\tau;l_g)=k\log(\tau)+\log\left[\,_1F_1\left(k, k+k_e, l_g\left(\lambda_e-4N_e\tau\right)\right)\right].
\label{ll_Lg2}
\end{equation} 

\noindent Where, recall, we assume $k=1.5$. Note that if we replace
$2N_e\tau$ with $t$, and drop constant terms, then we get an expression for the likelihood of
$t$ that does not depend on $N_e$, so our estimate of time in
generations does not depend on $N_e$.  

\begin{equation}
\ell(t;l_g)=k\log(t)+\log\left[\,_1F_1\left(k, k+k_e, l_g\left(\lambda_e-2t\right)\right)\right].
\label{ll_Lg3}
\end{equation} 

\noindent Finally, note that the rate 
at which recombination events occur on the
branch connecting the two shared haplotypes is $4N_e\tau$. We 
assume that the first such event marks the end of the
haplotype. However, there is a non-zero probability that a
recombination event occurring on this branch does not change the MRCA
of $a$ and $b$. Simulations suggest that for large numbers of
chromosomes, this probability is extremely small (Figure S12)
 and so we assume it is 0. In practice, for small samples, this
might be a non-negligible effect. 

\subsection*{Approximate likelihood for singleton count}

Recall that the physical length of the shared haplotype is $L_p$
bases. We assume that we can find this exactly. Then assuming a
constant mutation rate $\mu$ per base per generation, the sum of the number of
singletons on the shared haplotypes, $S^*$ has a Poisson distribution
with parameter $\theta L_p \tau$, where $\theta=4N_e\mu$.

Now consider the distribution of singletons on the unshared
haplotypes. To approximate this distribution,  we make the 
following three assumptions: 1) There is no
recombination on the unshared haplotypes over the region. 2) No other
lineage coalesces with the shared haplotype before it is broken. 3) The
distribution of the time to first coalescence of the unshared haplotypes is
exponential with parameter $N$ (Recall that $N$ is the number of
sampled individuals). In fact the true distribution is a mixture of
exponentials but the approximation at least matches the
correct mean, $\frac{1}{N}$ \cite{blum2005}. The variance is too small
because of the first assumption, however. 

Consider one of the unshared haplotypes. Conditional on the time ($\tau_1$) at
which it first coalesces with any other haplotype, the number of
singleton mutations it carries is Poisson with
parameter $\theta L_p \tau_1$ and so, using the assumptions above, the
unconditional distribution is geometric (on $0, 1 \dots$) with parameter
$\frac{1}{1+\frac{\theta L_p}{2N}}$. Therefore the distribution of the number
of mutations on both unshared haplotypes, $\Delta_S$, is the sum of two geometric
distributions which is negative binomial with parameters 
$\left(2, \frac{\theta L_p}{\theta L_p+2N}\right)$. The
density of the total number of singletons, $S$ is the convolution of
these two densities
\begin{equation}
L(\tau; l_p, s)=\sum_{x=0}^s f_{Po}\left(x; \theta l_p \tau\right)
f_{NB}\left(s-x; \left(2,  \frac{\theta l_p}{\theta l_p+2N}\right)\right)
\label{ll_S}
\end{equation}
where $f_{Po}\left(x; \lambda\right)=\frac{\lambda^x e^{-\lambda}}{x!}$ is the density of a Poisson
distribution with parameter $\lambda$ and $f_{NB}\left(x;
  \left(n,p\right) \right)=\binom{x+n-1}{x}(1-p)^np^x$ is the density of a negative binomial
distribution with parameters $(n,p)$. As with the genetic length, we
can write this in terms of $t$, the haplotype age in generations,  

\begin{equation}
L(t; l_p, s)=\sum_{x=0}^s f_{Po}\left(x; 2\mu l_p t\right)
f_{NB}\left(s-x; \left(2,  \frac{\theta l_p}{\theta l_p+2N}\right)\right)
\label{ll_S2}
\end{equation}

In practice we assume $\mu$ is known and estimate $\theta$
separately for each individual, for each chromosome, by counting the
number of singletons, multiplying by the number of chromosomes in the
sample, and dividing by the chromosome length.  
Then for each pair, we use use the average of these values in Equation \ref{ll_S2}.
A more accurate approach would be to 
compute the likelihood as a double convolution over the distribution
of both haplotypes with different values for $\theta$. An extension
would be to estimate $\theta$ separately for different regions of the
genome. 

\subsection*{Approximate full likelihood}

We can now write the approximate log-likelihood for $t$ as the sum
of Equation \ref{ll_Lg3} and the log of Equation \ref{ll_S2}, assuming
that the recombination process is independent of the mutational process,
\begin{equation}
\ell(t; l_g, l_p, s)=\ell(t;l_g) +\log\left[L(t;l_p, s)\right].
\end{equation}
We maximise it numerically with respect to $t$ in order to find the maximum likelihood
estimate (MLE). It is possible for this likelihood to be bimodal,
in which case we might find a local but not global optimum. However, this
seems to be rare.

\subsection*{1000 Genomes Data}
The 1000 Genomes data was obtained from
\texttt{ftp://ftp.1000genomes.ebi.ac.uk/vol1/ftp/}. 
The phase 1 release sequence data is in
\texttt{phase1/analysis\_results/integrated\_call\_sets},
and the array data is in
\texttt{phase1/analysis\_results/supporting/omni\_haplotypes}. In
order to generate the ``clean'' sequence data, we removed any sites
that fell in the list of low complexity regions found in
\texttt{technical/working/20140224\_low\_complexity\_regions/hs37d5-LCRs.txt}. 
Functional annotations are in
\texttt{phase1/analysis\_results/functional\_annotation}. 
Detailed explanations of the annotations can be found there, but briefly the
classifications are as follows: 
\begin{itemize}
\item
Loss-of-function: Includes premature stop codons, and essential splice
site disruptions. 
\item
Coding: Variants in coding regions. 
\item
Functional noncoding: Including variants in noncoding RNAs, promoters,
enhancers and transcription factor binding sites. 
\item
Unannotated: Any variant not included in any of the above categories. 
\end{itemize}
We included haplotypes in more than one of these categories if
they contained multiple variants. 
 
\subsection*{Code}
All the code we used to run simulations and analyse the 1000 Genomes
data is available from www.github.com/mathii/f2.

\section*{Acknowledgments}
Part of this work was completed while
I.M. was a research fellow at the Simons Institute for the Theory
of Computing at UC Berkeley. We thank Stephan Schiffels for extensive discussion, and providing
MSMC results. We also thank Richard Durbin, Alexander Kim and David Reich for helpful
suggestions. 

\newpage
\bibliography{refs}

\newpage
\section*{Supplementary information}
\renewcommand{\figurename}{Figure S{\hspace{-1mm}}}
\renewcommand{\tablename}{Table S{\hspace{-1mm}}}
\setcounter{figure}{0}

\begin{figure}[!h]
\begin{center}
\includegraphics{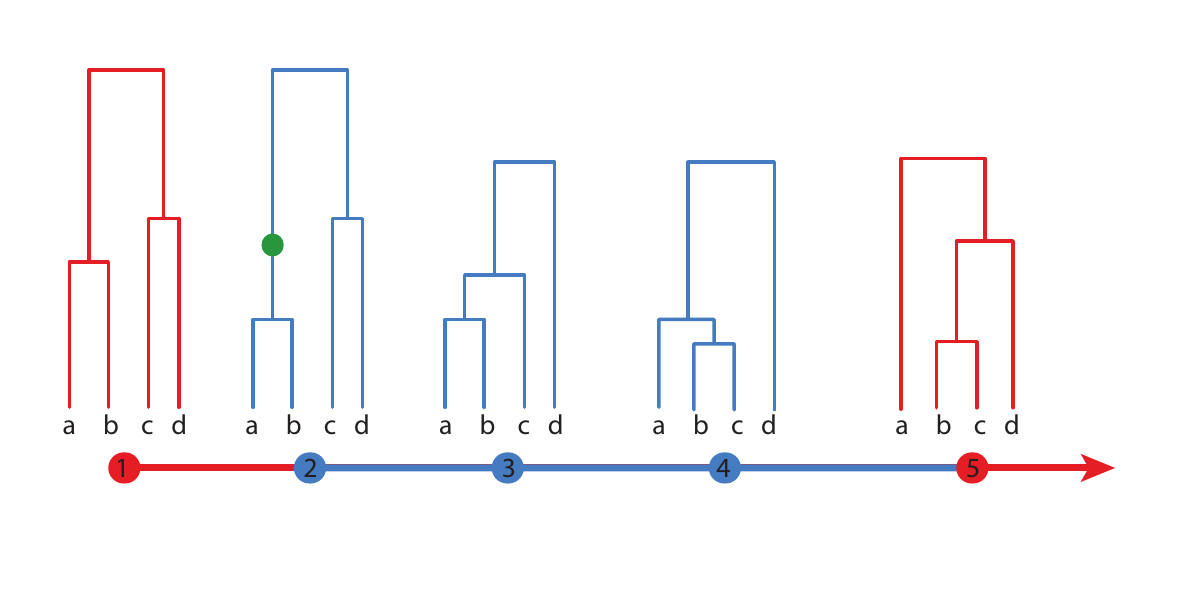}
\end{center}
\vskip-1cm
\caption{
\small {\bf{Example of an $f_2$ haplotype.}} Recombination events are shown
as a series of marginal trees, as we move left to right along a
sequence, so the tree above point $x$ is constant between $x$ and
$x+1$. In the blue region, $a$ and $b$ share an $f_2$ haplotype. At sites
2 and 3, $a$ and $b$ are unique nearest neighbours. A mutation at site
2 (green), will be detected as an $f_2$ variant. At site 4, they
are no longer unique nearest neighbours, but the TMRCA is
unchanged. At site 5, the TMRCA has changed and the haplotype
breaks. In the other direction, at site 1 the haplotype breaks because
the TMRCA of $a$ and $b$ changes, even though they are still unique
nearest neighbours. We count this as breaking the haplotype, even
though we cannot detect this event. 
}
\label{FigS1}
\end{figure}

\begin{figure}[!hp]
\begin{center}
\includegraphics[width=12cm]{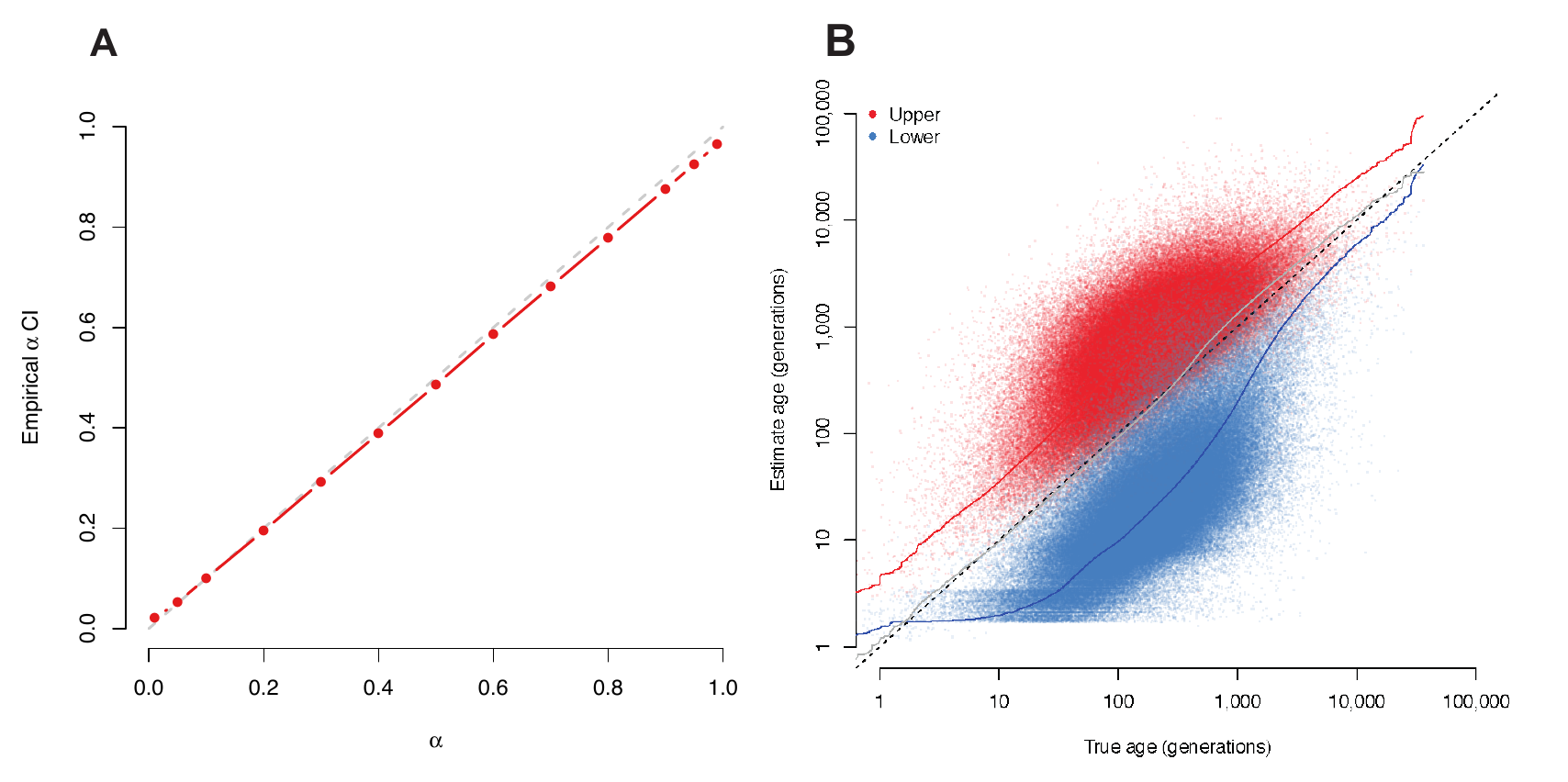}
\end{center}
\vskip-0cm
\caption{
\small
{\bf{Confidence intervals.}} 
{\bf{A}}: Coverage of approximate confidence intervals. We performed simulations
as described in Figure \ref{Fig2}, but only for chromosome
20. We computed approximate $\alpha$-confidence intervals using the
$\chi^2$ approximation to the distribution of the log-likelihood. This figure shows
the proportion of true haplotype ages that lie inside their
approximate confidence intervals. {\bf{B}}: Confidence intervals for
the simulations in Figure 2. For each
haplotype, we plot its true age against the upper and lower end of the
two-tailed 95\% confidence interval.
}
\label{FigS2}
\end{figure}

\begin{figure}
  \begin{center}
    \includegraphics[width=12cm]{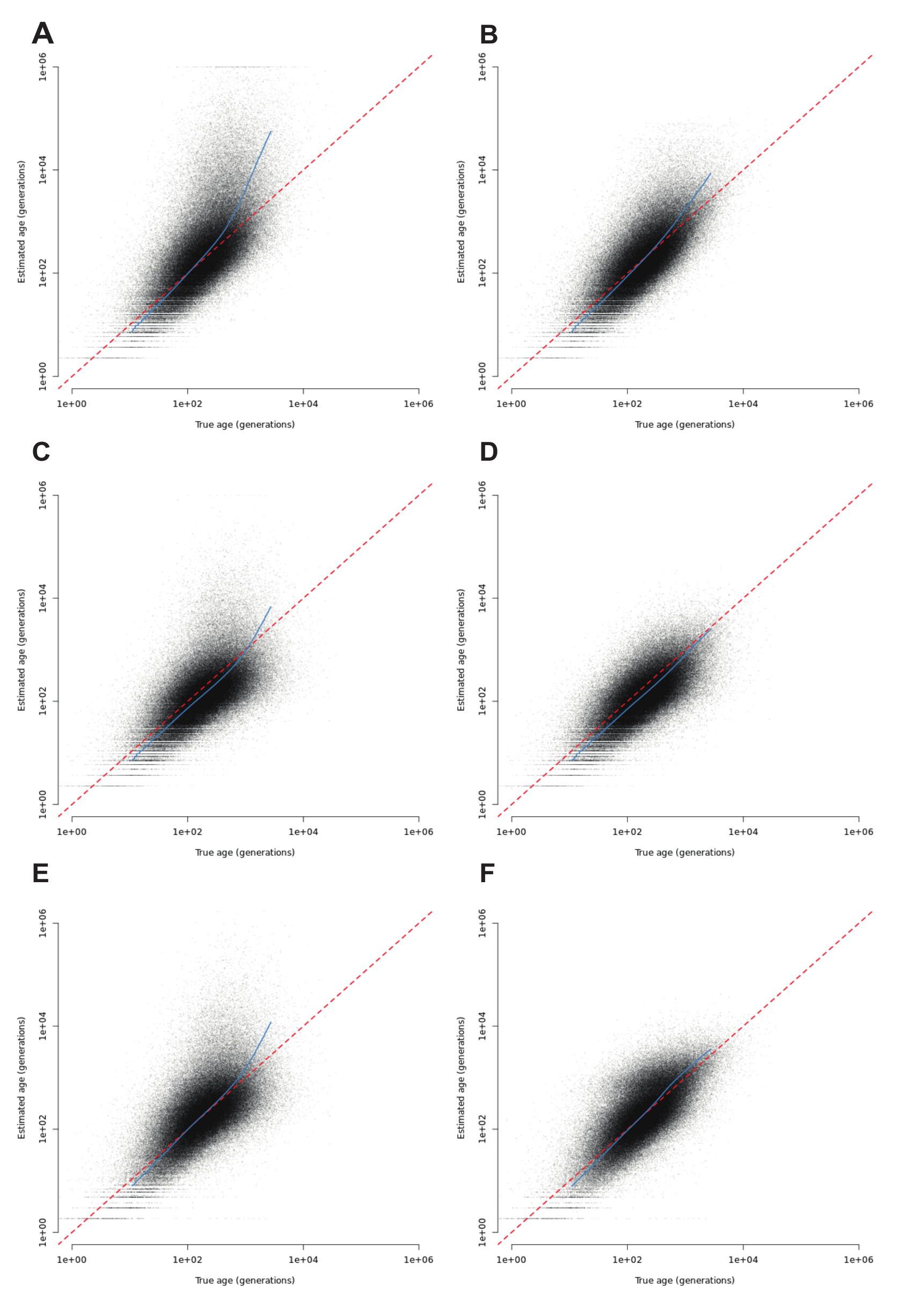}
  \end{center}
  \caption{
\small
{\bf{The effect of different terms of the likelihood.}} This
shows plots comparable to Figure \ref{Fig2}, based on
whole-genome simulations of 100 individuals. Grey dots show the
estimated age of each $f_2$ haplotype against the true age.
 The blue line is a qq plot of the distribution
of the MLEs (from 1\% to 99\% quantiles)
Each subfigure shows the result of using different information. {\bf{A}}, {\bf{C}} and
{\bf{E}} use just the genetic length and {\bf{B}}, {\bf{D}} and
{\bf{F}} use both the genetic length and the number of
singletons. {\bf{A}} and {\bf{B}} show the results if the true values
are used in the likelihood. {\bf{C}} and {\bf{D}} show the
corresponding results when the observed length is used without
accounting for the overestimate, {\bf{E}} and {\bf{F}} (the same as
Figure 2A) show the full
likelihood, including the correction to $L_g$.}
\label{FigS3}
\end{figure}

\begin{figure}[!hp]
  \vskip-10pt
  \begin{center}
    \includegraphics[]{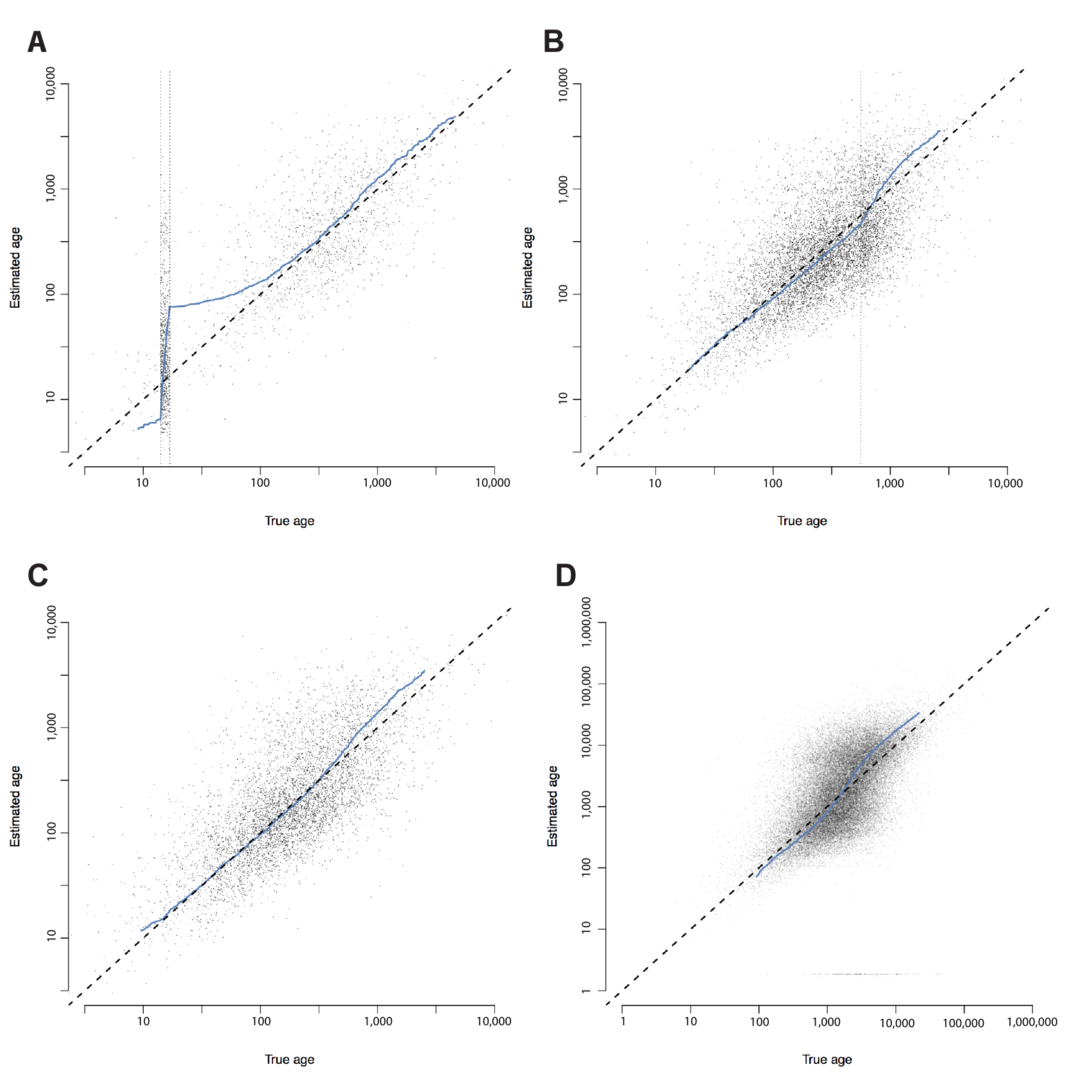}
  \end{center}

        \caption{
\small
{\bf{The effect of demography on the accuracy of
inference.}}
These plots were generated with the same parameters as
Figure \ref{Fig2} ($N_e=14,000$, $\mu=1.2\times 10^{-8}$), but
show only chromosome 20. Grey dots show the estimated age of each
detected $f_2$ haplotype against the true age (in generations). The
blue line is a qq plot of the distribution of the MLEs (from 1\% to 99\% quantiles).  We simulated different
demographic scenarios. {\bf{A}}: A bottleneck which reduces the
population by 99\% between 14 and 17 generations in the past (dotted
lines). {\bf{B}}: A population of size 14,000 which split into two isolated
populations, each of size 14,000, 560 generations ago (dotted line). {\bf{C}}: A population
growing exponentially by about 0.02\% per generation, to a present
size of 140,000. {\bf{D}}: A population where $N_e$ is actually 140,000 but 
we ran inference assuming it to be 14,000.
}
\vspace{3cm}
\label{FigS4}
\end{figure}

\begin{figure}[!hp]
  \vskip-30pt
        \centering
        \includegraphics[]{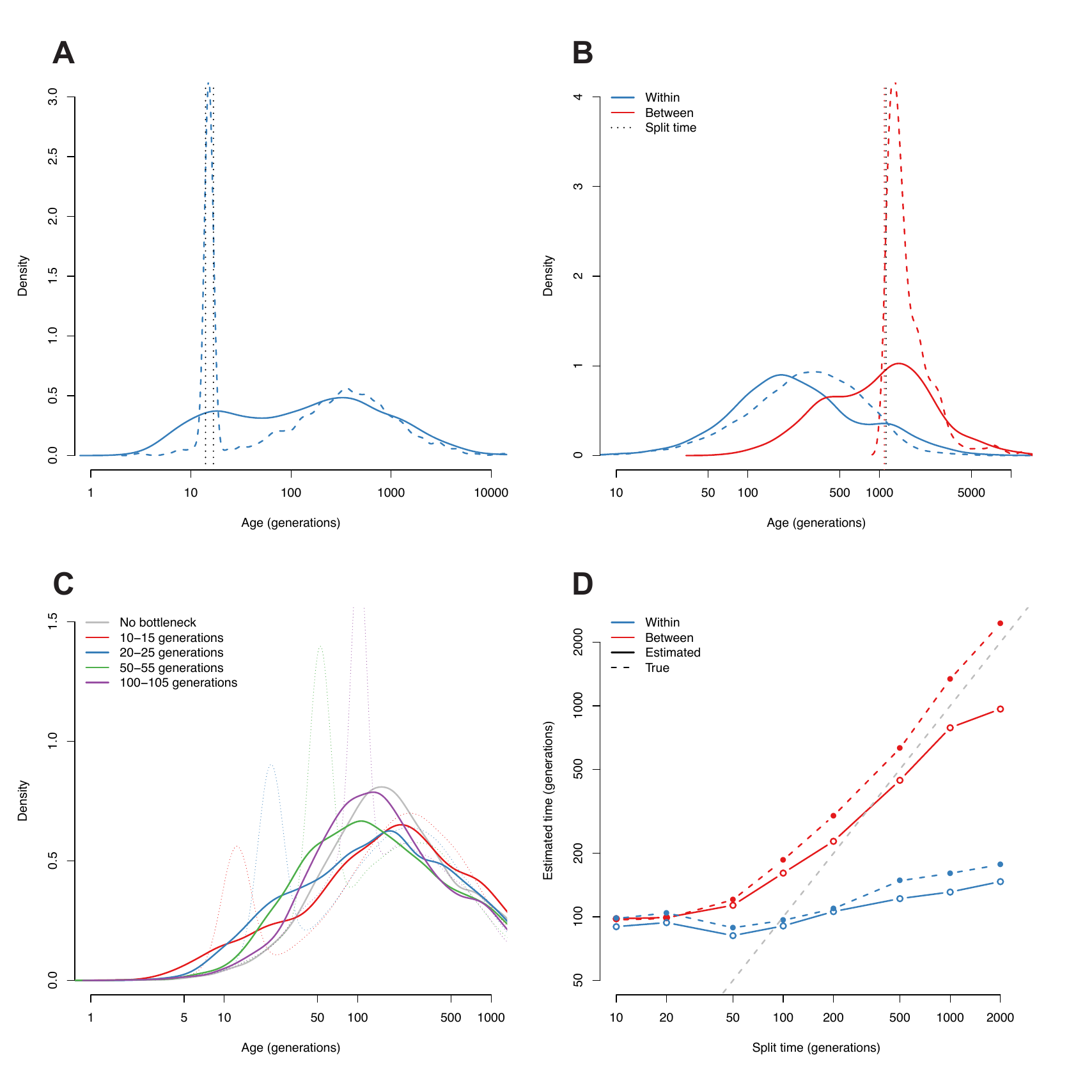}
  \caption{
\small
{\bf{The effect of demography on inferred age distributions.}}
Solid lines show the
density estimate of the distribution of the MLEs of the ages 
of detected $f_2$ haplotypes under different demographic scenarios. Dashed
lines show the true distributions. Parameters as in Figure \ref{Fig2}. {\bf{A}}: A
recent bottleneck.  Dotted lines show the time of the bottleneck
(population size reduced by 99\% for 3 generations).{\bf{B}}:
Population split 1120 generations ago with no subsequent migration. The
dotted line shows the time of the population split. The blue line
shows the estimated age of $f_2$ variants shared within a population and the red
line the estimated age of $f_2$ variants shared between populations. The red
dotted line shows the median of this distribution and the black dotted
line shows the split time. {\bf{C}}: Distributions for different
bottlenecks. As in {\bf{A}} but now reducing the population size by
90\% for 5 generations and using the recombination map from 
chromosome 10 instead of 20. {\bf{D}}: Median within and between population
ages, both estimated (solid line) and true (dashed line), for
different split times. As in {\bf{B}}, but again using the chromosome
10 recombination map
}
\label{FigS5}
\end{figure}
\vspace{5cm}

\begin{figure}[!hp]
\begin{center}
\includegraphics[width=\textwidth]{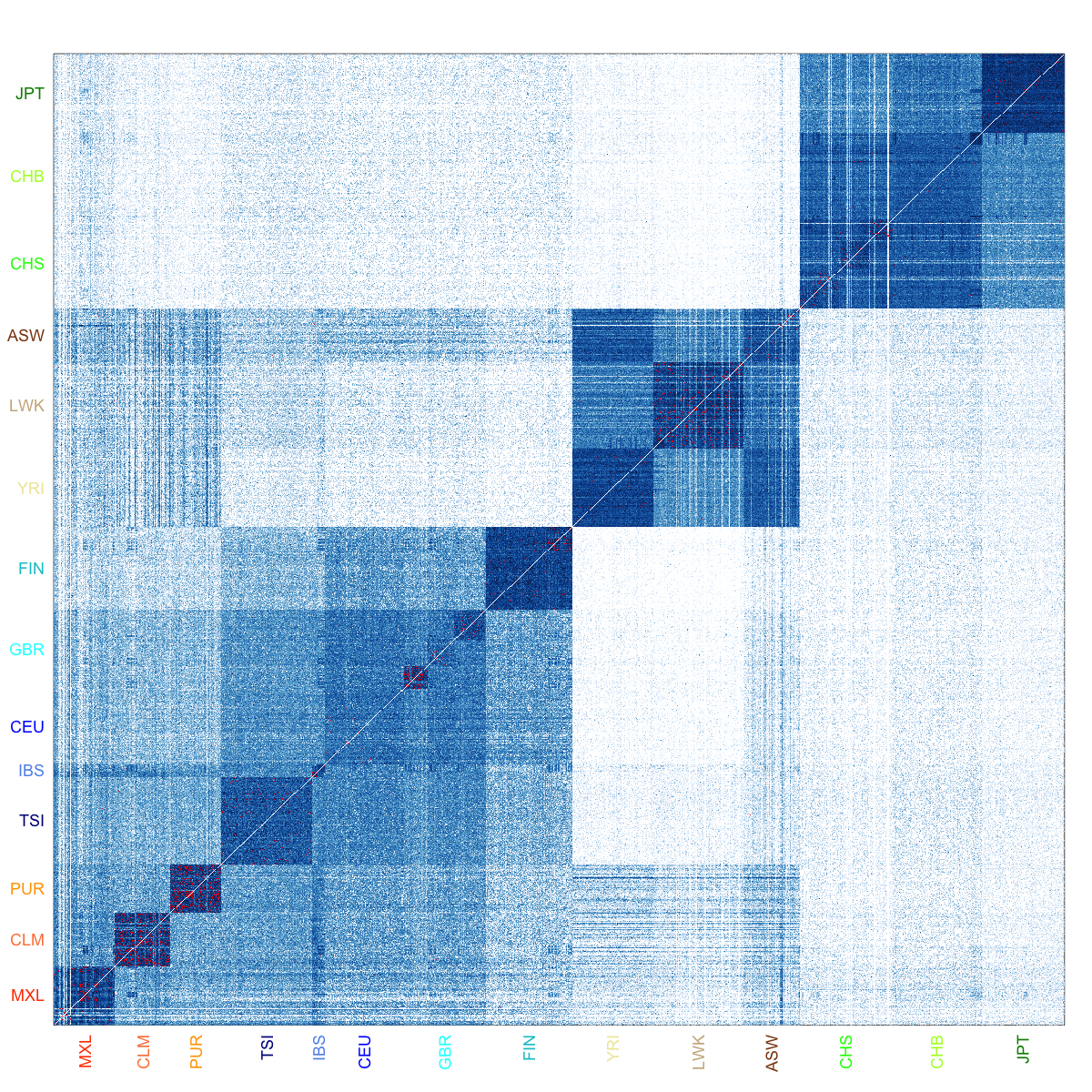}
\end{center}
\caption{
\small
{\bf{$f_2$ variant sharing across 1000 Genomes individuals.}} Colours from
white to blue to red show the number of $f_2$ variants shared between
each pair of individuals, normalised by the total in each
row. Individuals are ordered by populations, but only by sample name
within each population.
}
\label{FigS6}
\end{figure}

\begin{figure}[!hp]
\begin{center}
\includegraphics[]{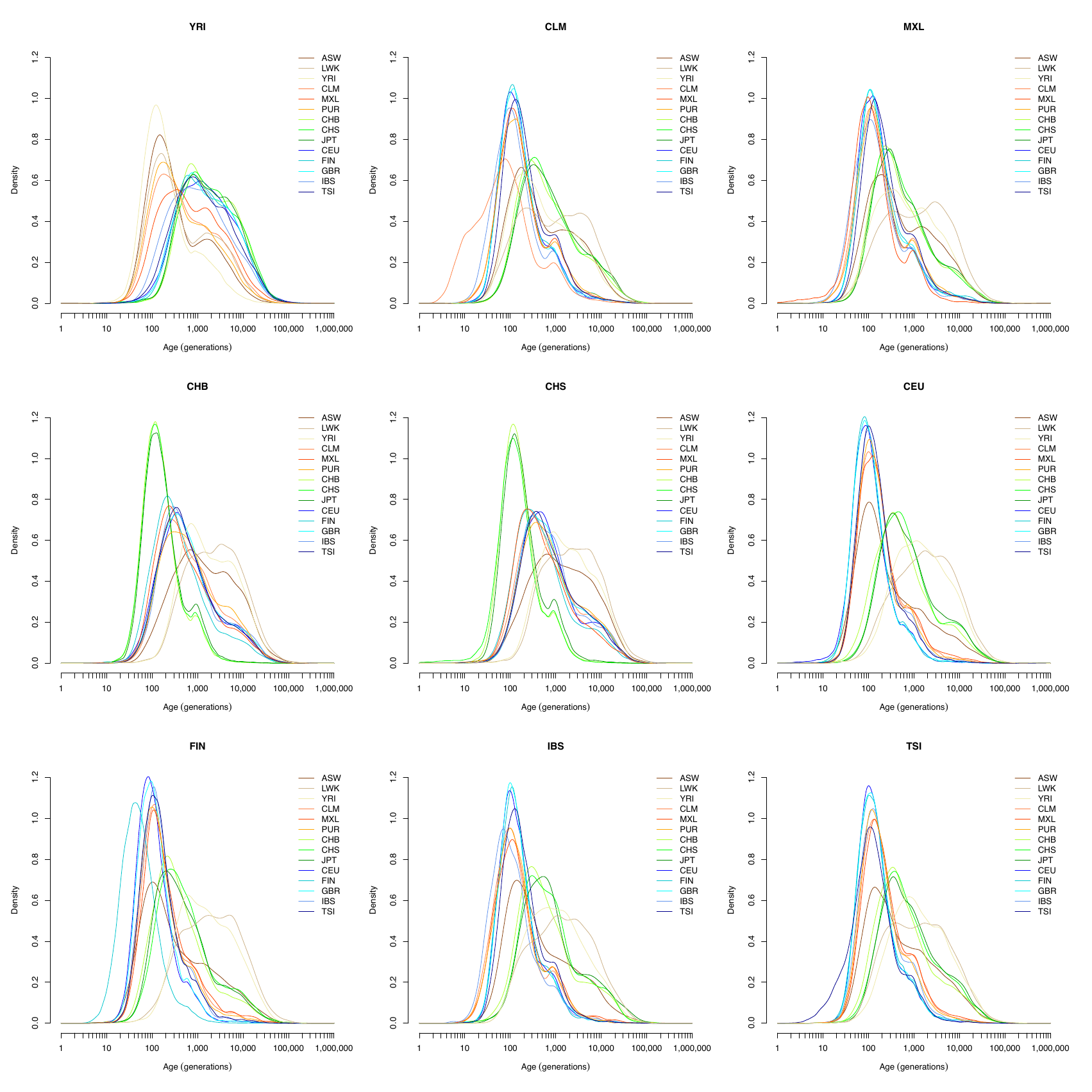}
\end{center}
\caption{
\small
 {\bf{1000 Genomes $f_2$ haplotype MLE age distributions.}} For
 the nine populations not included in
 Figure \ref{Fig3}. Each of these subfigures shows the distribution of
 ages of haplotypes shared between one population and each of the others. 
}
\label{FigS7}
\end{figure}

\begin{figure}[!hp]
\begin{center}
\includegraphics[]{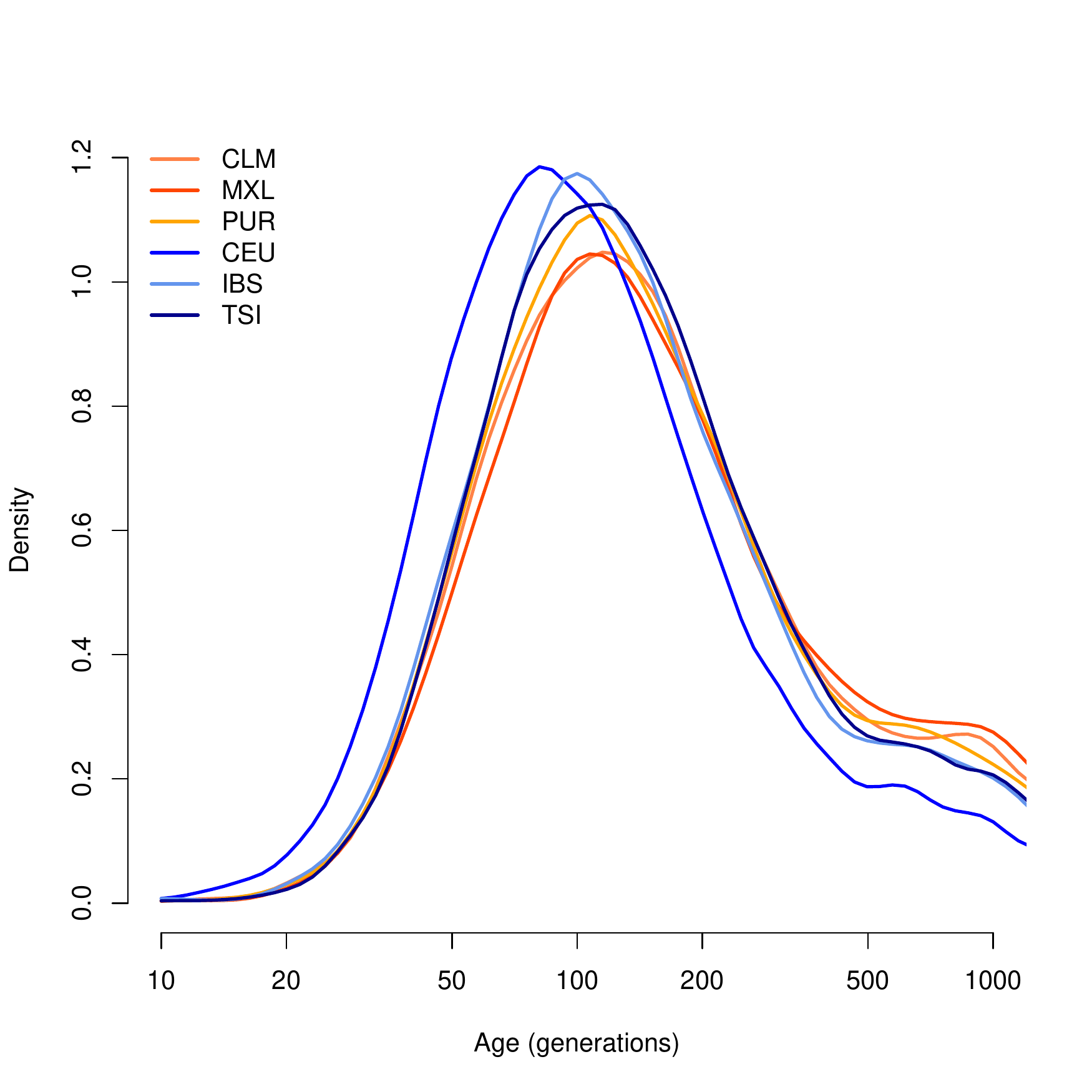}
\end{center}
\caption{
\small
  {\bf{GBR-American sharing.}} The density of the ages of $f_2$ haplotypes shared between GBR (UK) and
  each of CEU (NW European), CLM (Columbian), IBS (Spanish), MXL
  (Mexican), PUR (Puerto Rican) and TSI (Tuscan). See Table \ref{Tab1} for more
  complete descriptions of the populations.  
}
\label{FigS8}
\end{figure}

\begin{figure}[!hp]
  \vspace{4cm}
  \begin{center}
    \includegraphics[]{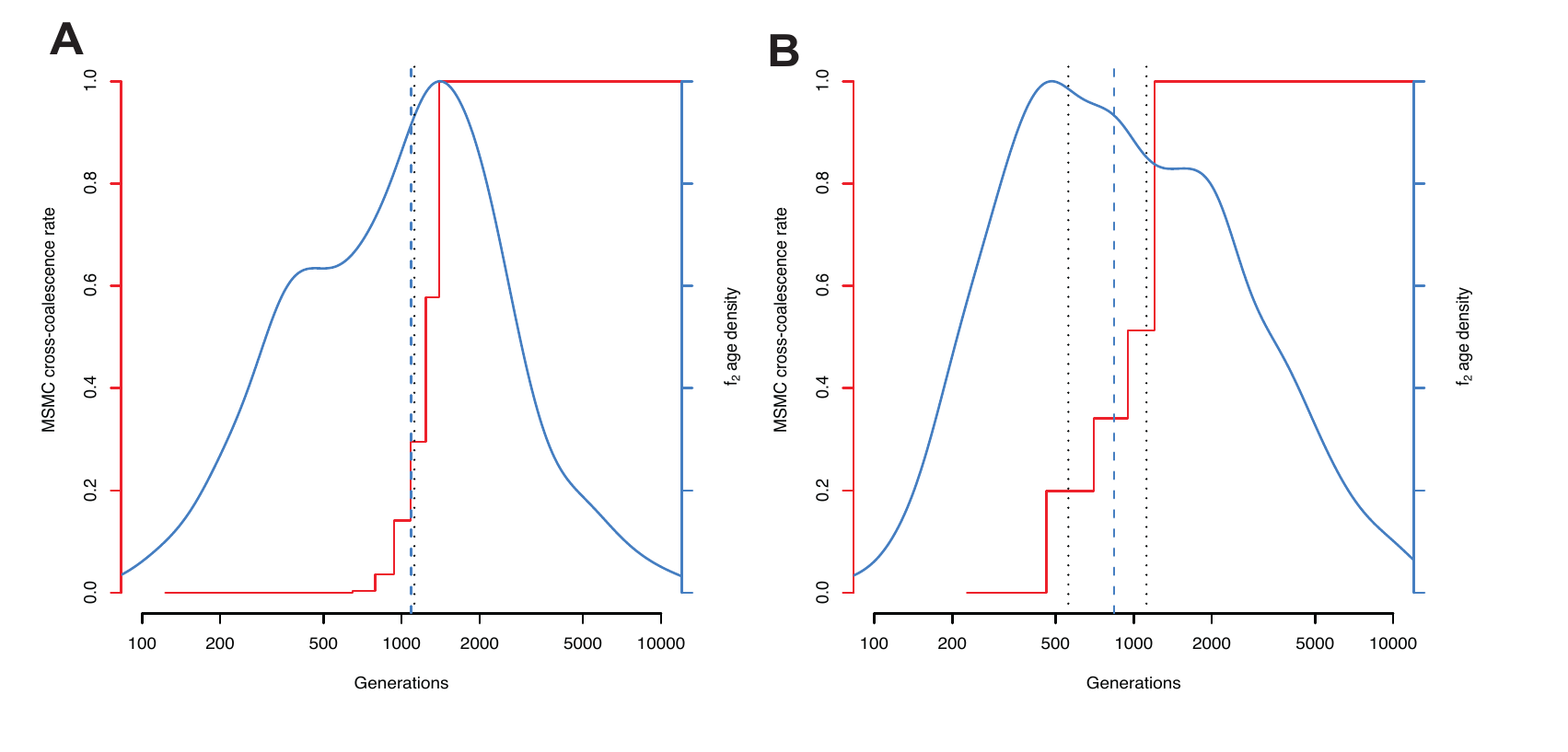}
  \end{center}
  \caption{
\small
  {\bf{The effect of post-split migration.}}
    Comparison of the distribution of the age of $f_2$ variants shared
  between populations (blue)
  and the gene flow estimated by MSMC with 4 haplotypes
  (red). $N_e=14,000$, $\mu=1.2\times 10^{-8}$, using the chromosome
  20 recombination map. In each case, the blue dashed line shows the
  median of the $f_2$ age distribution. {\bf{a}}: A scenario where the two populations split 1120
  generations ago (black dashed line). {\bf{b}}: A scenario where two populations split
  1120 generations ago, but there is migration at a rate of 1\% per
  year for 560 generations (black dashed lines at 560 and 1120
  generations). Note that in {\bf{b}}, the peak of the
  $f_2$ age density is shifted to the left relative to {\bf{a}},
  indicating that many of the $f_2$ variants shared between populations
  derive from post-split migration rather than predating the split. 
}
\vspace{4cm}
\label{FigS9}
\end{figure}

\begin{figure}
  \begin{center}
    \includegraphics[]{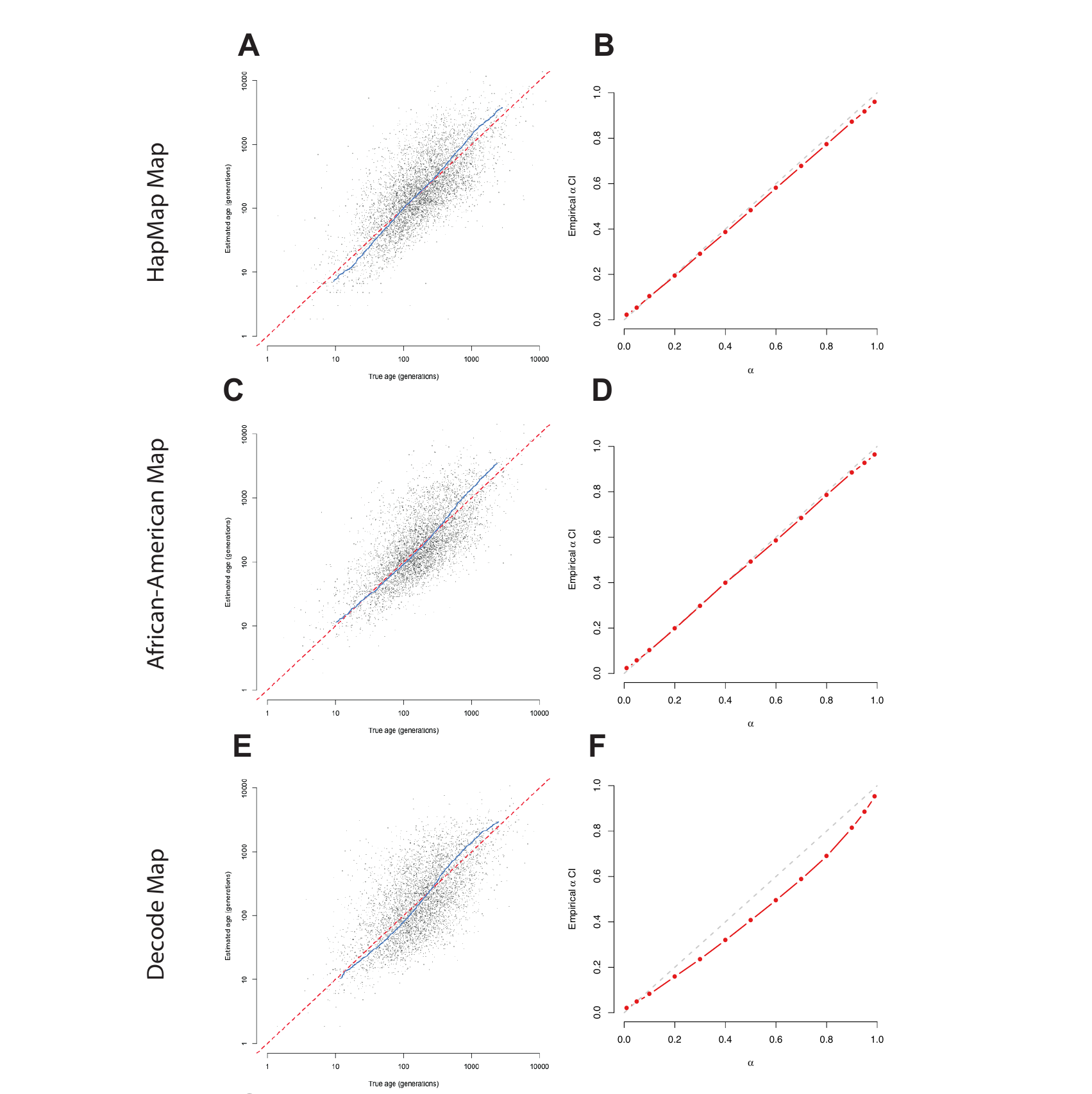}
  \end{center}

        \caption{This figure is continued on the next page.}
\end{figure}

\begin{figure}
  \ContinuedFloat
    \begin{center}
    \includegraphics[]{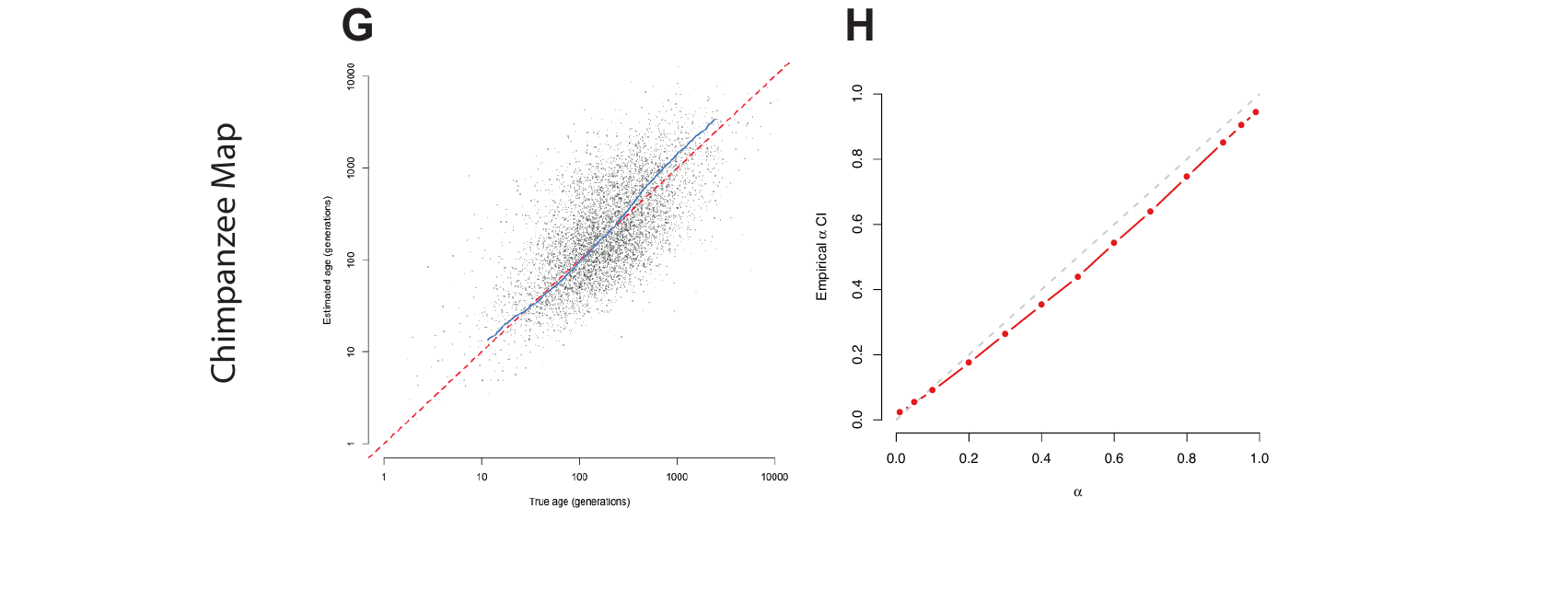}
  \end{center}
              \caption{\small{\bf{The effect of errors in the recombination
                    map.}}
                We simulated haplotypes with a
                different recombination map to the one used to determine genetic
                length. The left column shows true versus estimated
                ages for detected haplotypes, and a qq plot of the
                MLEs, as in Figure \ref{Fig2}A in the main text. The right
                column shows the coverage of the asymptotic
                confidence intervals as in Figure
                \ref{FigS2}A. Each row shows the results of simulations using a
                different map (references in main text), but in every case the HapMap combined
                map was used to determine the genetic length of the
                detected haplotypes: {\bf{A,B}}: Simulated using the HapMap
                map. {\bf{C,D}}: Simulated using a map derived from
                African Americans. {\bf{E,F}}: Simulated using a map
                derived from an Icelandic pedigree. {\bf{G,H}}:
                Simulated using a map derived from chimpanzees,
                rescaled to have the same total length as the human map. In
                each case, we simulated Chromosome 20 for 100
                individuals with
                $\mu=1.2\times10^{-8}$ per-base per-generation and 
              $N_e=14,000$.}
\vspace{5cm}
\label{FigS10}
\end{figure}

\begin{figure}[!hp]
\begin{center}
\includegraphics[]{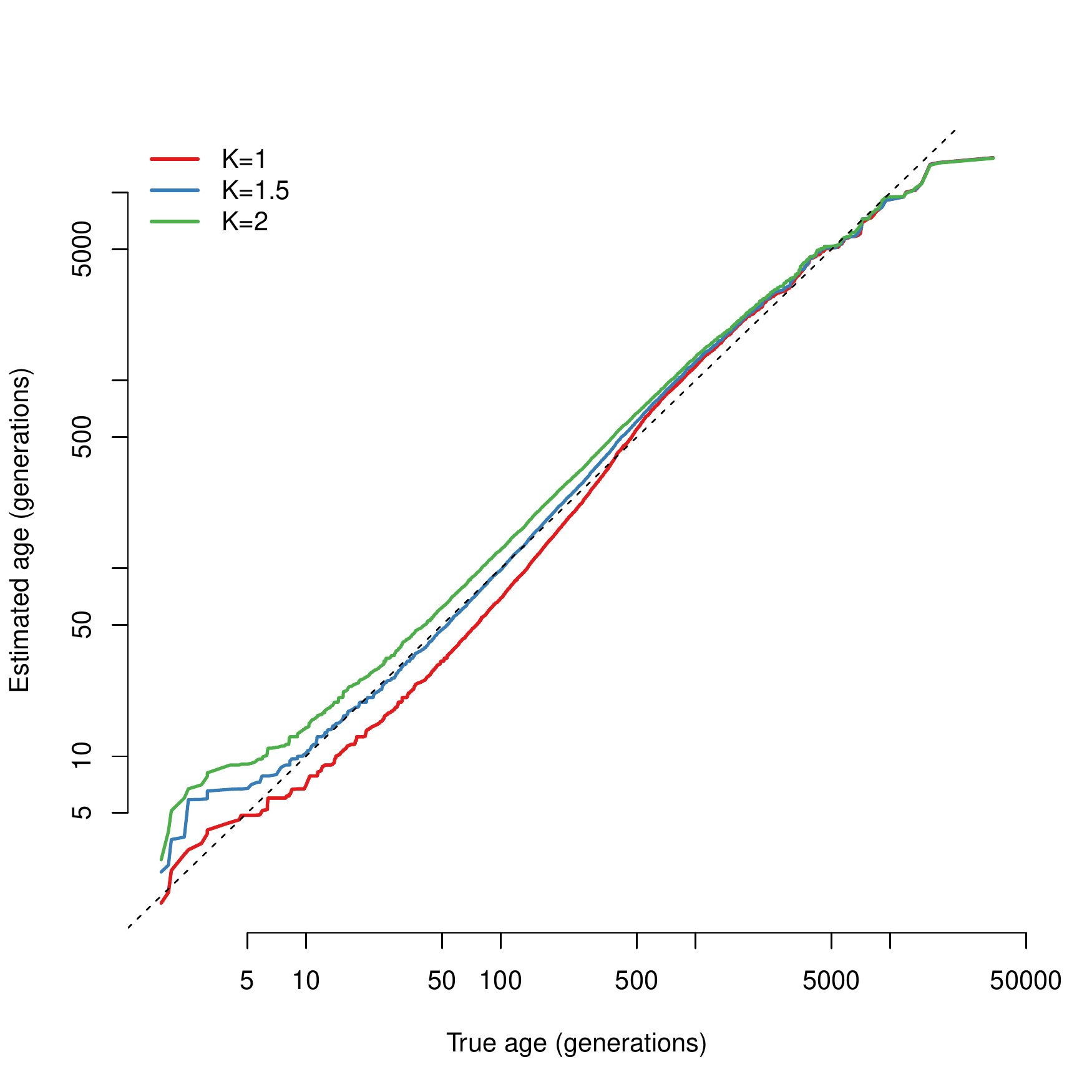}
\end{center}
\caption{
\small
{\bf{Effect of varying $k$.}}
This figure shows the effect on the density estimate of varying $k$, the shape parameter
of the gamma distribution used to model the genetic length of the
haplotypes. This shows qq plots generated from simulations as in Figure \ref{Fig2}, but for
chromosome 20 only, for $k=1$, 1.5 and 2. 
}
\label{FigS11}
\end{figure}

\begin{figure}[]
\begin{center}
\vskip-60pt
\includegraphics[]{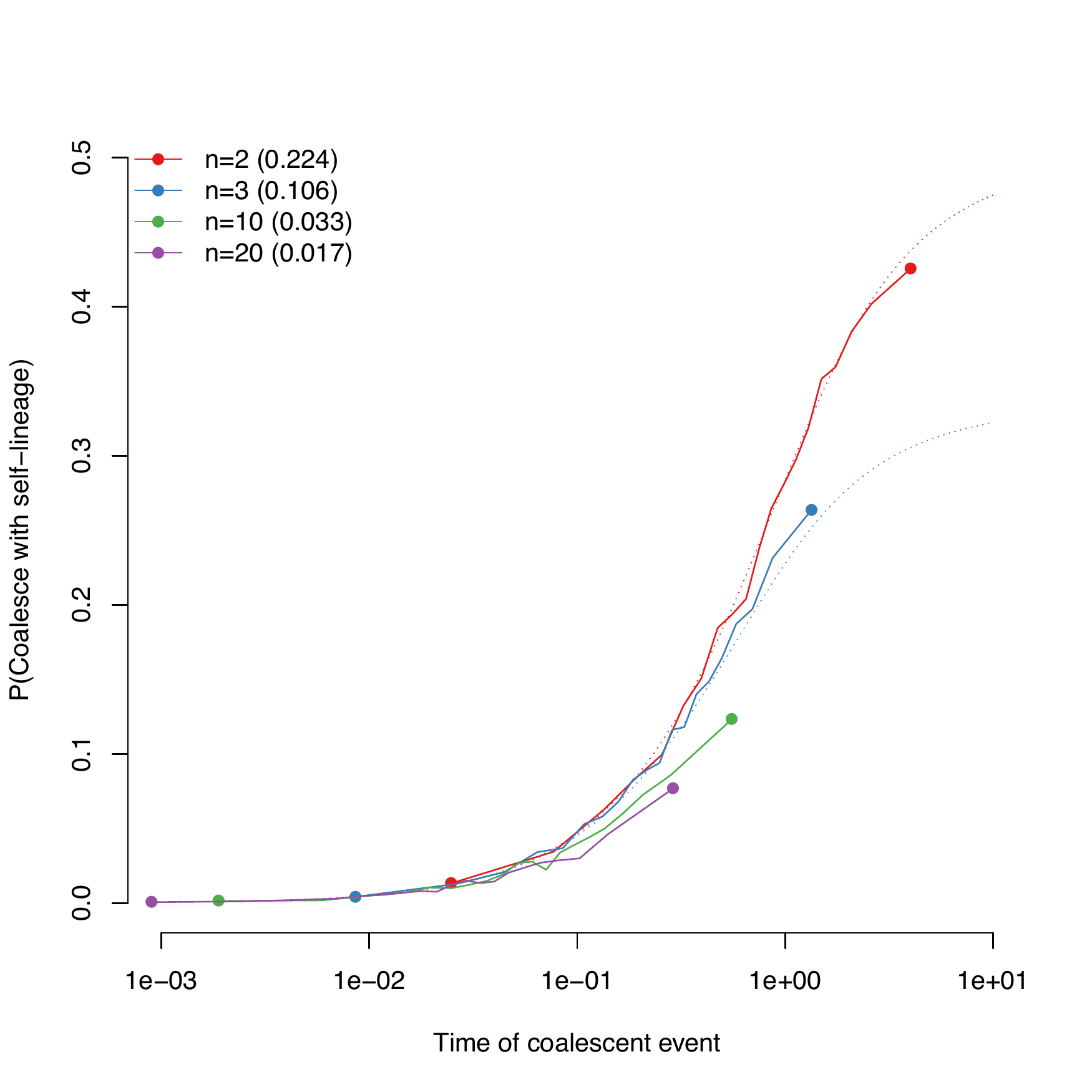}
\end{center}
\vskip-0.5cm
\caption{
\small
{\bf{Probability that recombinations do not change the TMRCA}}.
We used simulations to estimate the probability that the first recombination on the branch between two
haplotypes which are nearest neighbours 
does not change their TMRCA. For varying sample sizes
$n$, we simulated sequences with recombination using the SMC' algorithm. 
We find pairs of nearest neighbours, then count along the sequence
until the first recombination on the branch connecting them. The
plot shows the probability that this recombination does not change the
TMRCA ($\tau$) of those two samples, as a function of $\tau$. Solid
circles show 5\% and 95\% quantiles of the distribution of
coalescence times. The dashed
lines show the theoretical lower bounds on this probability; 
$\frac{1}{n}\left(1-\frac{1}{nt}\left(1-e^{-nt}\right)\right)$,
exact for $n=2$, which is achieved if there are no coalescences on
the tree except the one between the two nearest neighbour lineages.
The numbers in brackets in the legend show the
probability that a recombination does not change the TMRCA, averaged
over all events. Note that as $n$ increases, for fixed $\tau$, the
probability of not changing the TMRCA decreases, but in addition the
distribution of $\tau$ becomes smaller which also decreases the
probability of not changing the TMRCA.
}
\label{FigS12}
\end{figure}

\setcounter{table}{\arabic{figure}}

\begin{table}
\hspace{-2cm}
\tiny
\begin{tabular}{r|rrrrrrrrrrrrrr}
 & ASW & LWK & YRI & CLM & MXL & PUR & CHB & CHS & JPT & CEU & FIN & GBR & IBS & TSI\\
\hline
ASW & {\bf{ 32,995}} &  68,828 &  96,214 &  11,175 &  10,027 &  11,486 &   3,700 &   3,730 &   3,244 &   7,794 &   4,944 &   8,412 &   1,595 &   8,748\\
LWK &  68,828 & {\bf{135,521}} &  88,596 &  11,871 &   9,591 &  12,288 &   4,946 &   5,044 &   4,675 &   4,330 &   3,085 &   4,541 &   1,403 &   8,905\\
YRI &  96,214 &  88,596 & {\bf{101,317}} &  11,216 &   8,044 &  12,253 &   3,421 &   3,514 &   3,256 &   2,578 &   1,932 &   2,838 &     795 &   4,375\\
CLM &  11,175 &  11,871 &  11,216 & {\bf{ 22,080}} &  17,268 &  12,516 &   2,888 &   2,594 &   2,581 &  13,483 &   8,007 &  15,120 &   4,992 &  19,860\\
MXL &  10,027 &   9,591 &   8,044 &  17,268 & {\bf{ 30,772}} &  12,397 &   4,359 &   4,091 &   3,981 &  12,559 &   7,685 &  13,560 &   4,196 &  18,973\\
PUR &  11,486 &  12,288 &  12,253 &  12,516 &  12,397 & {\bf{ 15,732}} &   2,062 &   1,891 &   1,807 &  12,742 &   7,065 &  13,674 &   3,777 &  19,367\\
CHB &   3,700 &   4,946 &   3,421 &   2,888 &   4,359 &   2,062 & {\bf{ 60,480}} & 113,761 &  65,147 &   2,886 &   3,401 &   3,003 &     664 &   4,667\\
CHS &   3,730 &   5,044 &   3,514 &   2,594 &   4,091 &   1,891 & 113,761 & {\bf{ 69,260}} &  52,616 &   2,509 &   2,559 &   2,812 &     579 &   3,960\\
JPT &   3,244 &   4,675 &   3,256 &   2,581 &   3,981 &   1,807 &  65,147 &  52,616 & {\bf{109,214}} &   2,446 &   2,965 &   2,744 &     581 &   3,672\\
CEU &   7,794 &   4,330 &   2,578 &  13,483 &  12,559 &  12,742 &   2,886 &   2,509 &   2,446 & {\bf{ 21,517}} &  24,260 &  41,902 &   5,089 &  36,671\\
FIN &   4,944 &   3,085 &   1,932 &   8,007 &   7,685 &   7,065 &   3,401 &   2,559 &   2,965 &  24,260 & {\bf{ 51,084}} &  24,475 &   3,084 &  19,161\\
GBR &   8,412 &   4,541 &   2,838 &  15,120 &  13,560 &  13,674 &   3,003 &   2,812 &   2,744 &  41,902 &  24,475 & {\bf{ 27,314}} &   6,344 &  35,757\\
IBS &   1,595 &   1,403 &     795 &   4,992 &   4,196 &   3,777 &     664 &     579 &     581 &   5,089 &   3,084 &   6,344 & {\bf{  1,502}} &   6,931\\
TSI &   8,748 &   8,905 &   4,375 &  19,860 &  18,973 &  19,367 &   4,667 &   3,960 &   3,672 &  36,671 &  19,161 &  35,757 &   6,931 & {\bf{ 43,068}}\\
\end{tabular}
\caption{
\small
{\bf{1000 Genomes $f_2$ haplotype total counts.}} Counts of $f_2$ haplotypes shared between each pair of
  populations.}
\label{TabS13} 
\end{table}

\begin{table}
\hspace{-2cm}
\small
\begin{tabular}{r|rrrrrrrrrrrrrr}
 & ASW & LWK & YRI & CLM & MXL & PUR & CHB & CHS & JPT & CEU & FIN & GBR & IBS & TSI\\
\hline
ASW & {\bf{18.03}} & 11.63 & 17.92 &  3.05 &  2.49 &  3.42 &  0.63 &  0.61 &  0.60 &  1.50 &  0.87 &  1.55 &  1.87 &  1.46\\
LWK & 11.63 & {\bf{29.11}} & 10.38 &  2.04 &  1.50 &  2.30 &  0.53 &  0.52 &  0.54 &  0.53 &  0.34 &  0.53 &  1.03 &  0.94\\
YRI & 17.92 & 10.38 & {\bf{26.47}} &  2.12 &  1.38 &  2.53 &  0.40 &  0.40 &  0.42 &  0.34 &  0.24 &  0.36 &  0.65 &  0.51\\
CLM &  3.05 &  2.04 &  2.12 & {\bf{12.47}} &  4.36 &  3.79 &  0.50 &  0.43 &  0.48 &  2.64 &  1.43 &  2.83 &  5.94 &  3.38\\
MXL &  2.49 &  1.50 &  1.38 &  4.36 & {\bf{14.35}} &  3.42 &  0.68 &  0.62 &  0.68 &  2.24 &  1.25 &  2.31 &  4.54 &  2.93\\
PUR &  3.42 &  2.30 &  2.53 &  3.79 &  3.42 & {\bf{10.59}} &  0.39 &  0.34 &  0.37 &  2.73 &  1.38 &  2.79 &  4.91 &  3.59\\
CHB &  0.63 &  0.53 &  0.40 &  0.50 &  0.68 &  0.39 & {\bf{12.99}} & 11.73 &  7.55 &  0.35 &  0.38 &  0.35 &  0.49 &  0.49\\
CHS &  0.61 &  0.52 &  0.40 &  0.43 &  0.62 &  0.34 & 11.73 & {\bf{13.99}} &  5.91 &  0.30 &  0.28 &  0.32 &  0.41 &  0.40\\
JPT &  0.60 &  0.54 &  0.42 &  0.48 &  0.68 &  0.37 &  7.55 &  5.91 & {\bf{27.89}} &  0.32 &  0.36 &  0.35 &  0.47 &  0.42\\
CEU &  1.50 &  0.53 &  0.34 &  2.64 &  2.24 &  2.73 &  0.35 &  0.30 &  0.32 & {\bf{ 6.03}} &  3.07 &  5.54 &  4.28 &  4.40\\
FIN &  0.87 &  0.34 &  0.24 &  1.43 &  1.25 &  1.38 &  0.38 &  0.28 &  0.36 &  3.07 & {\bf{11.94}} &  2.96 &  2.37 &  2.10\\
GBR &  1.55 &  0.53 &  0.36 &  2.83 &  2.31 &  2.79 &  0.35 &  0.32 &  0.35 &  5.54 &  2.96 & {\bf{ 6.97}} &  5.09 &  4.10\\
IBS &  1.87 &  1.03 &  0.65 &  5.94 &  4.54 &  4.91 &  0.49 &  0.41 &  0.47 &  4.28 &  2.37 &  5.09 & {\bf{16.51}} &  5.05\\
TSI &  1.46 &  0.94 &  0.51 &  3.38 &  2.93 &  3.59 &  0.49 &  0.40 &  0.42 &  4.40 &  2.10 &  4.10 &  5.05 & {\bf{ 9.06}}\\
\end{tabular}
\caption{
\small
{\bf{1000 Genomes $f_2$ haplotype mean counts.}} Mean number of $f_2$ haplotypes shared between each pair of
  individuals, for each pair of populations.
} 
\label{TabS14} 
\end{table}

\begin{table}
\hspace{-1cm}
\small
\begin{tabular}{r|rrrrrrrrrrrrrr}
 & ASW & LWK & YRI & CLM & MXL & PUR & CHB & CHS & JPT & CEU & FIN & GBR & IBS & TSI\\
\hline
ASW & {\bf{  337}} &   527 &   247 &   385 &   406 &   371 & 1,306 & 1,272 & 1,503 &   184 &   232 &   190 &   311 &   358\\
LWK &   527 & {\bf{  317}} &   308 &   897 & 1,178 &   781 & 2,374 & 2,295 & 2,227 & 1,795 & 2,013 & 1,833 & 1,377 & 1,224\\
YRI &   247 &   308 & {\bf{  168}} &   421 &   651 &   351 & 1,640 & 1,710 & 1,649 & 1,449 & 1,453 & 1,394 &   996 & 1,209\\
CLM &   385 &   897 &   421 & {\bf{   70}} &   148 &   152 &   536 &   555 &   606 &   143 &   144 &   143 &   117 &   179\\
MXL &   406 & 1,178 &   651 &   148 & {\bf{  115}} &   156 &   402 &   432 &   441 &   148 &   159 &   150 &   131 &   191\\
PUR &   371 &   781 &   351 &   152 &   156 & {\bf{   38}} &   583 &   629 &   620 &   135 &   135 &   137 &   126 &   166\\
CHB & 1,306 & 2,374 & 1,640 &   536 &   402 &   583 & {\bf{  137}} &   140 &   149 &   481 &   321 &   499 &   496 &   483\\
CHS & 1,272 & 2,295 & 1,710 &   555 &   432 &   629 &   140 & {\bf{  134}} &   158 &   581 &   421 &   600 &   612 &   606\\
JPT & 1,503 & 2,227 & 1,649 &   606 &   441 &   620 &   149 &   158 & {\bf{  110}} &   615 &   380 &   621 &   650 &   667\\
CEU &   184 & 1,795 & 1,449 &   143 &   148 &   135 &   481 &   581 &   615 & {\bf{   98}} &    99 &   100 &   129 &   130\\
FIN &   232 & 2,013 & 1,453 &   144 &   159 &   135 &   321 &   421 &   380 &    99 & {\bf{   48}} &   106 &   131 &   132\\
GBR &   190 & 1,833 & 1,394 &   143 &   150 &   137 &   499 &   600 &   621 &   100 &   106 & {\bf{   90}} &   128 &   133\\
IBS &   311 & 1,377 &   996 &   117 &   131 &   126 &   496 &   612 &   650 &   129 &   131 &   128 & {\bf{   97}} &   158\\
TSI &   358 & 1,224 & 1,209 &   179 &   191 &   166 &   483 &   606 &   667 &   130 &   132 &   133 &   158 & {\bf{  120}}\\
\end{tabular}
\caption{\small
{\bf{1000 Genomes $f_2$ haplotype medians.}} Median estimated age (in generations) of the MLE of the $f_2$ haplotypes shared
  between each pair of populations, using array data to
  estimate the haplotypes.} 
\label{TabS15} 
\end{table}

\begin{table}
\hspace{-1cm}
\begin{tabular}{r|rrrrrrrrrrrrrr}
 & ASW & LWK & YRI & CLM & MXL & PUR & CHB & CHS & JPT & CEU & FIN & GBR & IBS & TSI\\
\hline
ASW & {\bf{ 43}} &  56 &  54 &  57 &  58 &  57 & 127 & 114 & 154 &  36 &  38 &  37 &  53 &  55\\
LWK &  56 & {\bf{ 21}} &  55 &  65 &  83 &  62 & 266 & 267 & 264 & 146 & 183 & 165 & 116 & 116\\
YRI &  54 &  55 & {\bf{ 44}} &  63 &  75 &  59 & 255 & 238 & 241 & 173 & 197 & 182 & 115 & 147\\
CLM &  57 &  65 &  63 & {\bf{  7}} &  35 &  34 &  94 &  92 &  95 &  41 &  41 &  42 &  27 &  49\\
MXL &  58 &  83 &  75 &  35 & {\bf{ 25}} &  37 &  78 &  83 &  82 &  43 &  44 &  43 &  29 &  51\\
PUR &  57 &  62 &  59 &  34 &  37 & {\bf{  6}} &  89 &  98 &  95 &  41 &  40 &  42 &  29 &  47\\
CHB & 127 & 266 & 255 &  94 &  78 &  89 & {\bf{ 44}} &  44 &  47 &  78 &  64 &  86 &  90 &  87\\
CHS & 114 & 267 & 238 &  92 &  83 &  98 &  44 & {\bf{ 34}} &  49 & 101 &  79 & 106 & 105 & 104\\
JPT & 154 & 264 & 241 &  95 &  82 &  95 &  47 &  49 & {\bf{ 32}} & 106 &  74 & 109 & 111 & 111\\
CEU &  36 & 146 & 173 &  41 &  43 &  41 &  78 & 101 & 106 & {\bf{ 29}} &  33 &  32 &  41 &  41\\
FIN &  38 & 183 & 197 &  41 &  44 &  40 &  64 &  79 &  74 &  33 & {\bf{ 13}} &  35 &  40 &  41\\
GBR &  37 & 165 & 182 &  42 &  43 &  42 &  86 & 106 & 109 &  32 &  35 & {\bf{ 15}} &  40 &  42\\
IBS &  53 & 116 & 115 &  27 &  29 &  29 &  90 & 105 & 111 &  41 &  40 &  40 & {\bf{ 24}} &  45\\
TSI &  55 & 116 & 147 &  49 &  51 &  47 &  87 & 104 & 111 &  41 &  41 &  42 &  45 & {\bf{ 18}}\\
\end{tabular}
\caption{
\small
{\bf{1000 Genomes $f_2$ haplotype 5\% quantiles.}} 5\% quantile of estimated age (in generations) of the MLE of the $f_2$ haplotypes shared
  between each pair of populations, using array data to
  estimate the haplotypes.} 
\label{TabS16} 
\end{table}

\begin{table}[]
\hspace{-1cm}
\tiny
\begin{tabular}{r|rrrrrrrrrrrrrr}
 & ASW & LWK & YRI & CLM & MXL & PUR & CHB & CHS & JPT & CEU & FIN & GBR & IBS & TSI\\
\hline
ASW & {\bf{ 7,962}} & 10,586 &  5,468 &  8,978 &  9,744 &  8,199 & 18,082 & 17,514 & 18,524 &  8,507 & 10,064 &  7,509 & 10,143 & 10,604\\
LWK & 10,586 & {\bf{ 8,542}} &  7,413 & 13,162 & 14,561 & 11,548 & 21,792 & 21,506 & 19,650 & 18,776 & 21,241 & 19,008 & 17,184 & 16,092\\
YRI &  5,468 &  7,413 & {\bf{ 2,871}} &  8,094 & 10,538 &  6,802 & 17,223 & 17,887 & 17,090 & 16,361 & 17,094 & 15,967 & 14,148 & 16,412\\
CLM &  8,978 & 13,162 &  8,094 & {\bf{ 1,238}} &  2,064 &  2,268 & 11,347 & 12,275 & 12,946 &  1,587 &  1,906 &  1,606 &  1,480 &  2,005\\
MXL &  9,744 & 14,561 & 10,538 &  2,064 & {\bf{ 1,291}} &  2,369 &  9,987 &  9,292 & 10,130 &  1,909 &  2,466 &  1,988 &  1,761 &  2,221\\
PUR &  8,199 & 11,548 &  6,802 &  2,268 &  2,369 & {\bf{ 1,039}} & 11,186 & 13,069 & 11,695 &  1,417 &  1,574 &  1,397 &  1,506 &  1,704\\
CHB & 18,082 & 21,792 & 17,223 & 11,347 &  9,987 & 11,186 & {\bf{ 1,084}} &  1,127 &  1,286 & 10,739 &  7,552 & 11,562 & 12,424 & 10,677\\
CHS & 17,514 & 21,506 & 17,887 & 12,275 &  9,292 & 13,069 &  1,127 & {\bf{ 1,108}} &  1,369 & 12,347 &  9,867 & 12,524 & 12,862 & 12,118\\
JPT & 18,524 & 19,650 & 17,090 & 12,946 & 10,130 & 11,695 &  1,286 &  1,369 & {\bf{ 1,204}} & 12,723 &  8,429 & 13,806 & 16,332 & 12,959\\
CEU &  8,507 & 18,776 & 16,361 &  1,587 &  1,909 &  1,417 & 10,739 & 12,347 & 12,723 & {\bf{   765}} &    741 &    780 &  1,177 &  1,117\\
FIN & 10,064 & 21,241 & 17,094 &  1,906 &  2,466 &  1,574 &  7,552 &  9,867 &  8,429 &    741 & {\bf{   296}} &    757 &  1,089 &  1,188\\
GBR &  7,509 & 19,008 & 15,967 &  1,606 &  1,988 &  1,397 & 11,562 & 12,524 & 13,806 &    780 &    757 & {\bf{   709}} &  1,174 &  1,158\\
IBS & 10,143 & 17,184 & 14,148 &  1,480 &  1,761 &  1,506 & 12,424 & 12,862 & 16,332 &  1,177 &  1,089 &  1,174 & {\bf{ 1,154}} &  1,397\\
TSI & 10,604 & 16,092 & 16,412 &  2,005 &  2,221 &  1,704 & 10,677 & 12,118 & 12,959 &  1,117 &  1,188 &  1,158 &  1,397 & {\bf{ 1,137}}\\
\end{tabular}
\caption{
\small
{\bf{1000 Genomes $f_2$ haplotype 95\% quantiles.}} 95\% quantile of estimated age (in generations) of the MLE of
  the$f_2$ haplotypes shared
  between each pair of populations, using array data to
  estimate the haplotypes.} 
\label{TabS17} 
\end{table}

\begin{table}[h!]
\begin{center}
\begin{tabular}{r|rr|rr|rr}
Chr & $k^{Array}_e$ & $\lambda^{Array}_e$ & $k^{Seq}_e$ & $\lambda^{Seq}_e$ &
$\theta_{Afr}$& $\theta_{N.Afr} $\\
\hline
1&0.413&60&0.386&222&3.47&2.18\\
2&0.539&186&0.377&260&3.79&2.45\\
3&0.580&234&0.389&278&3.88&2.44\\
4&0.576&215&0.382&276&3.94&2.40\\
5&0.564&213&0.374&263&3.80&2.44\\
6&0.570&228&0.386&282&3.76&2.39\\
7&0.588&222&0.380&278&3.62&2.38\\
8&0.533&195&0.378&275&4.24&2.57\\
9&0.468&78&0.406&264&3.48&2.03\\
10&0.557&195&0.393&253&3.90&2.41\\
11&0.561&236&0.369&273&3.88&2.52\\
12&0.585&215&0.398&259&3.74&2.35\\
13&0.528&178&0.389&250&3.35&2.00\\
14&0.507&156&0.390&273&3.23&2.03\\
15&0.350&46&0.373&203&3.20&1.96\\
16&0.619&213&0.430&308&3.74&2.41\\
17&0.506&138&0.388&235&3.66&2.31\\
18&0.616&222&0.416&277&3.94&2.44\\
19&0.652&218&0.431&291&3.41&2.28\\
20&0.617&226&0.407&252&3.73&2.46\\
21&0.611&162&0.428&191&2.90&1.82\\
22&0.526&138&0.421&258&2.32&1.60\\
\hline
\end{tabular}
\end{center}
\caption{
\small
{\bf{Estimated error parameters for 1000 Genomes data.}} We show
  error parameters $k_e$ and $\lambda_e$ for each chromosome estimated with both array and
  sequence data, and $\theta$ ($\times10^{-3}$) estimated by counting singletons, for
  each chromosome, showing the median value for non-African (N.Afr)
  and African (Afr) populations separately. Note that there is
  substantial variation in error parameters between chromosomes using
  array but not sequence data, suggesting that this is due to
  variations in the density of markers on the array, when we use array data to
  estimate the haplotypes.}
\label{TabS18} 
\end{table}

\end{document}